\documentclass[journal]{IEEEtran}
\usepackage[colorlinks,linkcolor=red,anchorcolor=red,citecolor=red,bookmarks=true]{hyperref}  
\usepackage[T1]{fontenc}
\ifCLASSINFOpdf
\usepackage[pdftex]{graphicx} 
\DeclareGraphicsExtensions{.pdf,.pdf,.jpeg,.png}
\else
\usepackage[dvips]{graphicx}
\fi
\usepackage[compress,nospace]{cite}
\usepackage[cmex10]{amsmath}
\usepackage{epstopdf,amsthm,stfloats,siunitx,amssymb,wasysym,algorithm,algorithmic,array,url, color,subfigure} 
\usepackage{footnote}
\usepackage{soul} 
\usepackage{color, xcolor} 
\usepackage{booktabs}
\usepackage{array, threeparttable}
\begin{document}
	
\title{URGLQ: An Efficient Covariance Matrix Reconstruction Method for Robust Adaptive Beamforming }
	
\author{Tao~Luo, Peng~Chen,~\IEEEmembership{Senior~Member,~IEEE}, Zhenxin~Cao,~\IEEEmembership{Member,~IEEE}, Le~Zheng,~\IEEEmembership{Senior~Member,~IEEE}, Zongxin~Wang,~\IEEEmembership{Member,~IEEE}

\thanks{T.~Luo is with the State Key Laboratory of Millimeter Waves, Southeast University, Nanjing 210096, China, and also with State Key Laboratory of Integrated Services Networks, Xidian University, Xi'an 710071, China (e-mail: luotaoseu@seu.edu.cn).}
\thanks{P.~Chen is with the State Key Laboratory of Millimeter Waves, Southeast University, Nanjing 210096, China, and also with State Key Laboratory of Integrated Services Networks, Xidian University, Xi'an 710071, China (e-mail: chenpengseu@seu.edu.cn).}
\thanks{Z.~Cao is with the State Key Laboratory of Millimeter Waves, Southeast University, Nanjing 210096, China (e-mail: caozx@seu.edu.cn).}
\thanks{L.~Zheng is with the School of Information and Electronics, Beijing Institute of Technology, Beijing 100081, China, and also with the Beijing Institute of Technology Chongqing Innovation Center, Chongqing 401120, China  (e-mail: le.zheng.cn@gmail.com).}
\thanks{Z.~Wang is with the State Key Laboratory of Millimeter Waves, Southeast University, Nanjing 210096, China (e-mail: wangzx@seu.edu.cn).}
\thanks{
This work was supported in part by 
the Natural Science Foundation for Excellent Young Scholars of Jiangsu Province (Grant No. BK20220128),
the National Key R\&D Program of China (Grant No. 2019YFE0120700), the Open Fund of ISN State Key Lab (Grant No. ISN24-04), the Industry-University-Research Cooperation Foundation of The Eighth Research Institute of China Aerospace Science and Technology Corporation (Grant No. SAST2021-039), and the National Natural Science Foundation of China (Grant No. 61801112).
}
\thanks{\textit{(Corresponding author: Peng Chen)}}
}

	\markboth{IEEE Transactions on Aerospace and Electronic Systems}%
	{Shell \MakeLowercase{\textit{et al.}}: Bare Demo of IEEEtran.cls for Journals}
	
	\maketitle
	
\begin{abstract} 
    The computational complexity of the conventional adaptive beamformer is relatively large, and the performance degrades significantly due to the model mismatch errors and the unwanted signals in received data. In this paper, an efficient unwanted signal removal and Gauss-Legendre quadrature (URGLQ)-based covariance matrix reconstruction method is proposed. Different from the prior covariance matrix reconstruction methods, a projection matrix is constructed to remove the unwanted signal from the received data, which improves the reconstruction accuracy of the covariance matrix. Considering that the computational complexity of most matrix reconstruction algorithms is relatively large due to the integral operation, we proposed a Gauss-Legendre quadrature-based method to approximate the integral operation while maintaining accuracy. Moreover, to improve the robustness of the beamformer, the mismatch in the desired steering vector is corrected by maximizing the output power of the beamformer under a constraint that the corrected steering vector cannot converge to any interference steering vector. Simulation results and prototype experiments demonstrate that the performance of the proposed beamformer outperforms the compared methods and is much closer to the optimal beamformer in different scenarios.
\end{abstract}
	
	\begin{IEEEkeywords}
		Covariance matrix reconstruction, desired signal removal, robust adaptive beamforming, Gauss-Legendre quadrature, steering vector estimation.
	\end{IEEEkeywords}
	\section{Introduction} \label{sec1}
	Adaptive beamforming is an array signal processing technology that has been widely applied in radar, sonar, wireless communication, and many other fields~\cite{zheng2017robust,van2004detection}. It is a data-based beamformer that adjusts the weights adaptive according to the received data to extract the desired signal and suppress the interference and noise~\cite{reed1974rapid}. The minimum variance distortionless response (MVDR) is one of the most well-known adaptive beamforming algorithms with the assumption that the desired signal and antenna array structure are known accurately~\cite{capon1969high}. MVDR has excellent resolution and interference suppression capability, but it is sensitive to the steering vector and covariance matrix mismatch. The effectiveness of the beamformer degrades severely, especially when the desired signal is presented in the received data, which is inevitable in practice~\cite{vorobyov2003robust}. Therefore, a lot of work has been spent on how to improve the robustness of the adaptive beamformer during the last decades~\cite{du2010fully,gu2012robust,yang2019robust}.
	
	Generally, the robust adaptive beamforming techniques can be divided into several types: diagonal loading technique,  eigenspace-based technique and covariance matrix reconstruction~\cite{9534656}. The diagonal loading is a widely used method, which adds a scaled identity matrix to the covariance matrix to get robustness~\cite{carlson1988covariance, 4020392}. However, how to choose the optimal diagonal loading factor is a difficult problem, and it decides the performance of the beamformer directly. An uncertainty set of the steering vector is proposed to calculate the diagonal loading factors precisely in~\cite{li2003robust}. A simple tridiagonal loading method called automatic tridiagonal loading is proposed to enhance the robustness in~\cite{zhang2019simple}. The eigenspace-based technique uses the orthogonality between the steering vector and subspace to correct the nominal steering vector and estimate the covariance matrix~\cite{jia2013robust,9612158,7929337,9775699}. However, it is hard to obtain an accuracy subspace and the signal subspace can be covered by the noise subspace when the SNR is low. An eigenvalue beamformer is proposed to resolve the unknown signal of interest whose spatial signature lies in a known subspace in~\cite{4490112}. Ref.~\cite{8516386} uses the subspace fitting and subspace orthogonality techniques to improve the performance of the beamformer.
	
	The covariance matrix reconstruction technique is a novel method, which separates the desired signal component away from the sample covariance matrix to enhance the robustness of the beamformer~\cite{7155471, mohammadzadeh2019robust}. In~\cite{gu2012robust}, the interference-plus-noise covariance matrix (IPNCM) is firstly reconstructed by using the Capon spectrum to integrate the nominal steering vector over an angle range that does not contain the desired signal direction. Based on this, an annulus uncertainty set is proposed to replace the normal linear integral interval to improve the robustness in~\cite{huang2015robust}, which gives the algorithm better performance but higher complexity. The covariance matrix reconstruction in~\cite{mohammadzadeh2020maximum} follows a similar method to~\cite{gu2012robust,huang2015robust}, but the maximum entropy power spectrum (MEPS) is used to replace the Capon spectrum to reconstruct the matrix, and the performance is further improved. Estimating the interference steering vectors and power is also widely used to reconstruct the covariance matrix. Ref.~\cite{yuan2017robust} proposes a matrix reconstruction method by searching the interference steering vectors inside the intersection of the interference subspace. The interference steering vector is estimated based on ad-hoc parameters in~\cite{zheng2018covariance}. Ref.~\cite{sun2021robust} introduces a matrix reconstruction method based on subspace and gradient vector. The adaptive beamforming method under the colored noise is discussed in~\cite{9517025}. The impacts of interference power estimation on robust adaptive beamforming are firstly analyzed in~\cite{zheng2019robust}, and a matrix reconstruction method via simplified interference power estimation is introduced then. Based on it, two novel IPNCM reconstruction methods by estimating the power and steering vectors of interference are proposed in~\cite{9508950}. To reduce the algorithm complexity, computational efficient matrix reconstruction algorithms are proposed in~\cite{ruan2013robust, ruan2016robust, s21237783}. A low computational complexity beamformer using the covariance matrix taper technique is proposed in~\cite{zhang2015interference}.
	
	In this paper, to further improve the beamformer performance and reduce the algorithm's computational complexity, an unwanted signal removal and Gauss-Legendre quadrature (URGLQ)-based covariance matrix reconstruction method is proposed. In most existing algorithms, the IPNCM is reconstructed based on the received data, and it contains the desired signals, which will affect the accuracy of the reconstruction. Thus, a projection matrix is constructed to remove the desired signal information from the received data. The quasi-matrix can be obtained by projecting the received data onto the projection matrix, which contains little desired information. The quasi-matrix is then used to reconstruct IPNCM based on the Capon spectrum. Due to that the quasi-matrix has less desired information than the received data, the reconstruction of the IPNCM-based method is more accurate. Considering that the conventional method to calculate the integral is to approximate it by polynomial summation, which usually leads to high computational complexity, a low-complexity algorithm based on 3-order Gauss-Legendre quadrature (GLQ) is introduced to simplify the integral operation. It reduces the computational complexity of the algorithm while maintaining high algebraic precision. Furthermore, the presumed desired signal steering vector is corrected by maximizing the beamformer output power, which is solved under the constraint that the corrected steering vector can not converge to any interference. By combining the reconstructed IPNCM and the corrected desired signal steering vector, the proposed adaptive beamformer can be obtained. Simulations and experiments are done to explore the performance of the algorithm. To summarize,
		we make the contributions as follows: 
	\begin{itemize}
		\item  \textbf{A new method for removing the desired signal from received signal:} A projection matrix is constructed to remove the desired signal information from the received data. By projecting the received data onto the projection matrix, a matrix that contains little desired information can be obtained.
		
		\item \textbf{A new IPNCM reconstruction method with 3-order Gauss-Legendre quadrature:} Gauss-Legendre quadrature is used to approximate the integral operation and reduce the computational complexity. And a new IPNCM reconstruction method based on the desired signal removal and GLQ is proposed.
		
		\item \textbf{A method for robust beamforming:} To improve the robustness and performance of the beamformer, a new beamforming method based on IPNCM and steering vector estimation is proposed. Simulations and experimental results show that the algorithm is effective.
	\end{itemize}

	The rest of the paper is organized as follows. The signal model for the adaptive beamforming technique is introduced in Section~\ref{secsignal}. The preliminary for IPNCM estimation is discussed in Section~\ref{secpre}. In Section~\ref{secprop}, the novel algorithm is proposed in detail. After numerical simulations and experiments, the performances of the proposed beamformer under different scenarios are demonstrated in Section~\ref{secsim}. The conclusion is presented in \mbox{Section~\ref{secconc}} finally.
	
	\emph{Notations:} Upper-case and lower-case boldface letters denote the matrices and column vectors, respectively. $(\cdot)^{\mathrm{T}}$ denotes matrix transpose, while the Hermitian transpose is denoted as $(\cdot)^{\mathrm{H}}$. $\mathcal{E}\left\{\cdot\right\}$ stands for the expectation operator of stochastic variables. $\left \| \cdot \right \|_{2} $ denotes the $\ell_{2}$ norm. $\operatorname{Tr}\left \{ \cdot \right \} $ denotes the trace of a matrix, and it is equal to the sum of the diagonal elements of the matrix.
	
	\section{Signal Model for Adaptive Beamforming} \label{secsignal}
\begin{figure}
		\centering 
		\includegraphics[width=0.4\textwidth]{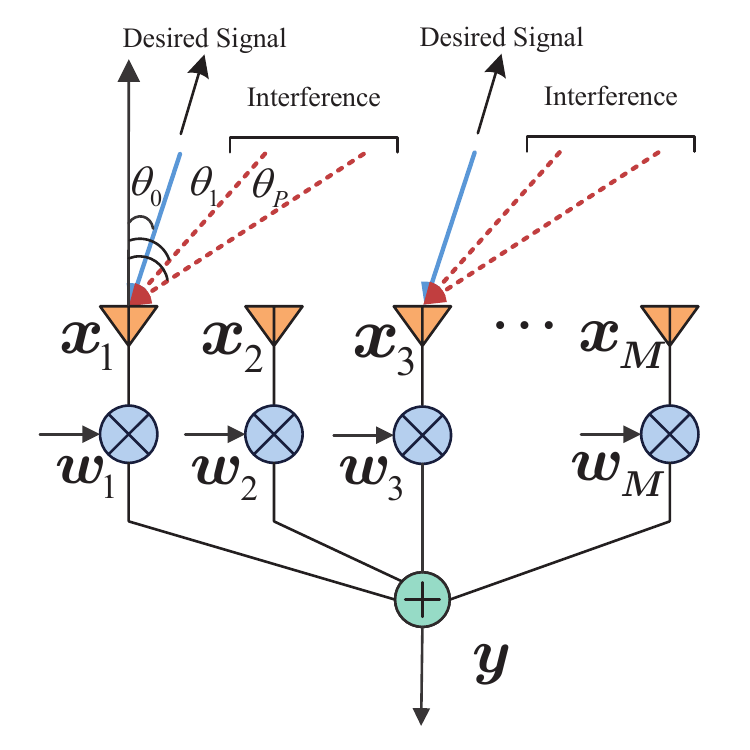}
		\caption{Uniform linear array for adaptive beamforming.}
		\label{ULA}
	\end{figure}

	Without loss of generality, a uniform linear array (ULA) with $M$ sensors is considered in this paper, as shown in Fig.~\ref{ULA}. The ULA receives $P+1$ far-field narrow-band signals composing of $1$ desired signal $s_0(t)$ and $P$ interference signals $s_p(t)$ ($p=1,2,\dots, P$). The received signal sampled at the $k$-th snapshot can be expressed as
	\begin{align}
		\boldsymbol{x}(k)& \triangleq [x_{1}(k),x_{2}(k),\dots,x_{M}(k)]^{\mathrm{T}}\\
		& = \underbrace{\boldsymbol{a}(\theta_{0})s_{0}(k)}_{\boldsymbol{x}_{\mathrm{s}}(k)}+\underbrace{\sum\limits_{p=1}^{P}\boldsymbol{a}(\theta_{p})s_{p}(k)}_{\boldsymbol{x}_{\mathrm{i}}(k)}+\boldsymbol{x}_{\mathrm{n}}(k),\notag
	\end{align}
	where $\boldsymbol{x}_{\mathrm{s}}(k) \in \mathbb{C}^{M\times 1}$, $\boldsymbol{x}_{\mathrm{i}}(k) \in \mathbb{C}^{M\times 1}$ and $\boldsymbol{x}_{\mathrm{n}}(k) \in \mathbb{C}^{M\times 1}$ are the desired signal, interference and noise, respectively, and these signals are statistically independent. $x_{m}(k)$ $(m = 1,2,\dots,M)$ represents the received data at the $m$-th sensor. $\theta_{0}$ denotes the direction of the desired signal, and $\theta_{p}$ is that of the $p$-th interference. $\boldsymbol{a}(\theta)$ defines a steering vector and can be given as 
	\begin{equation}
		\begin{aligned}
			\boldsymbol{a}(\theta) &= [a_{1},a_{2},\cdots,a_{M}]\\
			&=[1,e^{j\frac{2\pi d}{\lambda }\sin \theta },\dots,e^{j\frac{2(M-1)\pi d}{\lambda }\sin \theta}]^{\mathrm{T}},
		\end{aligned}
	\end{equation}
	where $d$ is the spacing between the adjacent sensors, and $\lambda $ denotes the wavelength.
	
	An adaptive beamforming method can be adopted to eliminate the interference, with a beamforming weight being $\boldsymbol{w}=[w_{1},w_{2},...,w_{M}]^{\mathrm{T}} \in \mathbb{C}^{M\times 1}$, the beamforming output can be obtained as
	\begin{equation}
		{y}(k)= \boldsymbol{w}^{\mathrm{H}}\boldsymbol{x}(k).
	\end{equation}
	Traditionally, the optimal weight $\boldsymbol{w}_{\mathrm{opt}}$ is obtained by maximizing the output signal-to-interference-plus-noise ratio (SINR)
	\begin{equation}
		\begin{aligned}
			\label{sinr}
			\mathrm{SINR} 
			&= \frac{\mathcal{E}\left \{\boldsymbol{w}^{\mathrm{H}}\boldsymbol{x}_{\mathrm{s}}\boldsymbol{x}^{\mathrm{H}}_{\mathrm{s}}\boldsymbol{w}\right \}}{\mathcal{E}\left\{\boldsymbol{w}^{\mathrm{H}}(\boldsymbol{x}_{\mathrm{i}}+\boldsymbol{x}_{\mathrm{n}})(\boldsymbol{x}_{\mathrm{i}}+\boldsymbol{x}_{\mathrm{n}})^{\mathrm{H}}\boldsymbol{w}\right\}}\\
			&= \frac{\sigma_{\mathrm{s}}^{2}\left | \boldsymbol{w}^{\mathrm{H}}\boldsymbol{a}(\theta_{0}) \right |^{2}}{\boldsymbol{w}^{\mathrm{H}}\boldsymbol{R}_{\mathrm{INF}}\boldsymbol{w}},
		\end{aligned}
	\end{equation}
	where $\sigma_{\mathrm{s}}^{2} = \mathcal{E}\left \{ |s_{0}(k)|^2\right \} $ denotes the power of the desired signal and $\boldsymbol{R}_{\mathrm{INF}}\in \mathbb{C}^{M \times M}$ is the IPNCM. When the interference and noise are irrelevant, $\boldsymbol{R}_{\mathrm{INF}}$ can be expressed as  
	\begin{equation}
		\begin{aligned}
			\boldsymbol{R}_{\mathrm{INF}} 
			&=\sum_{p=1}^{P} \sigma_{p}^{2}\boldsymbol{a}(\theta_{p})\boldsymbol{a}^{\mathrm{H}}(\theta_{p}) + \sigma_{\mathrm{n}}^{2} \boldsymbol{\mathrm{I}}_{M},
		\end{aligned}
	\end{equation}
	where $\sigma_{p}^{2}$ is the $p$-th interference power, $\sigma_{\mathrm{n}}^{2}$ is the noise power, and $\boldsymbol{\mathrm{I}}_{M}$ is an identity matrix with the size $M \times M$.
	
	The problem of maximizing (\ref{sinr}) is mathematically equivalent to the following minimum variance distortionless respond (MVDR) beamforming problem
	\begin{equation}
		\label{beamforming probelm}
		\min_{\boldsymbol{w}}\boldsymbol{w}^{\mathrm{H}}\boldsymbol{R}_{\mathrm{INF}}\boldsymbol{w}  \quad \mathrm{s.t.} \,\boldsymbol{w}^{H}\boldsymbol{a}(\theta_{0})=1. 
	\end{equation}
	The objective function in (\ref{beamforming probelm}) is to minimize the power of the interference and noise, and the constraint ensures that the power of the desired signal is not affected. The solution to (\ref{beamforming probelm}) as the MVDR weight is
	\begin{equation}
		\begin{aligned}
			\boldsymbol{w}_{\mathrm{opt}} =\frac{\boldsymbol{R}_{\mathrm{INF}}^{-1}\boldsymbol{a}(\theta_{0})}{\boldsymbol{a}^{\mathrm{H}}        (\theta_{0})\boldsymbol{R}_{\mathrm{INF}}^{-1}\boldsymbol{a}(\theta_{0})}.
			\label{mvdr}
		\end{aligned}
	\end{equation}
	
	In the practical system, since the received signal $\boldsymbol{x}(k)$ contains both the desired signal and the interference, the matrix $\boldsymbol{R}_{\mathrm{INF}}$ cannot be estimated accurately from $\boldsymbol{x}(k)$. Hence, we replace $\boldsymbol{R}_{\mathrm{INF}}$ by the following sample covariance matrix (SCM) 
	\begin{equation}
		\label{R_scm}
		\hat{\boldsymbol{R}}=\frac{1}{K} \sum_{k=1}^{K}\boldsymbol{x}(k)\boldsymbol{x}^{\mathrm{H}}(k),
	\end{equation}
	where $K$ is the number of snapshots. Then, substitute (\ref{R_scm}) into (\ref{mvdr}), and the MVDR beamformer turns to the sample matrix inversion (SMI) beamformer \cite{reed1974rapid}
	\begin{equation}
		\boldsymbol{w}_{\mathrm{SMI}}=\frac{\hat{\boldsymbol{R}}^{-1}\boldsymbol{a}(\theta_{0})}{\boldsymbol{a}^{\mathrm{H}}(\theta_{0})\hat{\boldsymbol{R}}^{-1}\boldsymbol{a}(\theta_{0})}. 
		\label{mvdr beamformer}
	\end{equation}
	
	However, there is a large gap between the SCM $\hat{\boldsymbol{R}}$ and the IPNCM $\boldsymbol{R}_\mathrm{INF}$, and the steering vector errors in the practical array degrade the performance of the beamformer $\boldsymbol{w}$ \cite{9452163}. Thus, in this paper, we are trying to estimate both the IPNCM $\boldsymbol{R}_\mathrm{INF}$ and the desired signal steering vector $\boldsymbol{a}(\theta_{0})$ accurately from the received signal $\boldsymbol{x}(k)$ to improve the robustness and the output SINR of the adaptive beamforming method.

	
	
	\section{Preliminary for IPNCM Estimation}\label{secpre}
	To estimate the IPNCM precisely, an efficient reconstruction method based on the Capon spatial spectrum has been proposed in \cite{zhu2020robust}, where the IPNCM is estimated by integrating over an angle range separated from the desired signal direction \cite{zhu2019covariance}. 
	
	First, the Capon spatial spectrum is obtained by substituting the MVDR beamformer (\ref{mvdr beamformer}) back into the objective function of (\ref{beamforming probelm})
	\begin{equation}
		\label{capon}
		P(\theta)=\frac{1}{\boldsymbol{a}^{\mathrm{H}}(\theta)\hat{\boldsymbol{R}}^{-1}\boldsymbol{a}(\theta)}. 
	\end{equation}
	
	\begin{figure}
		\centering 
		\includegraphics[width=0.3\textwidth]{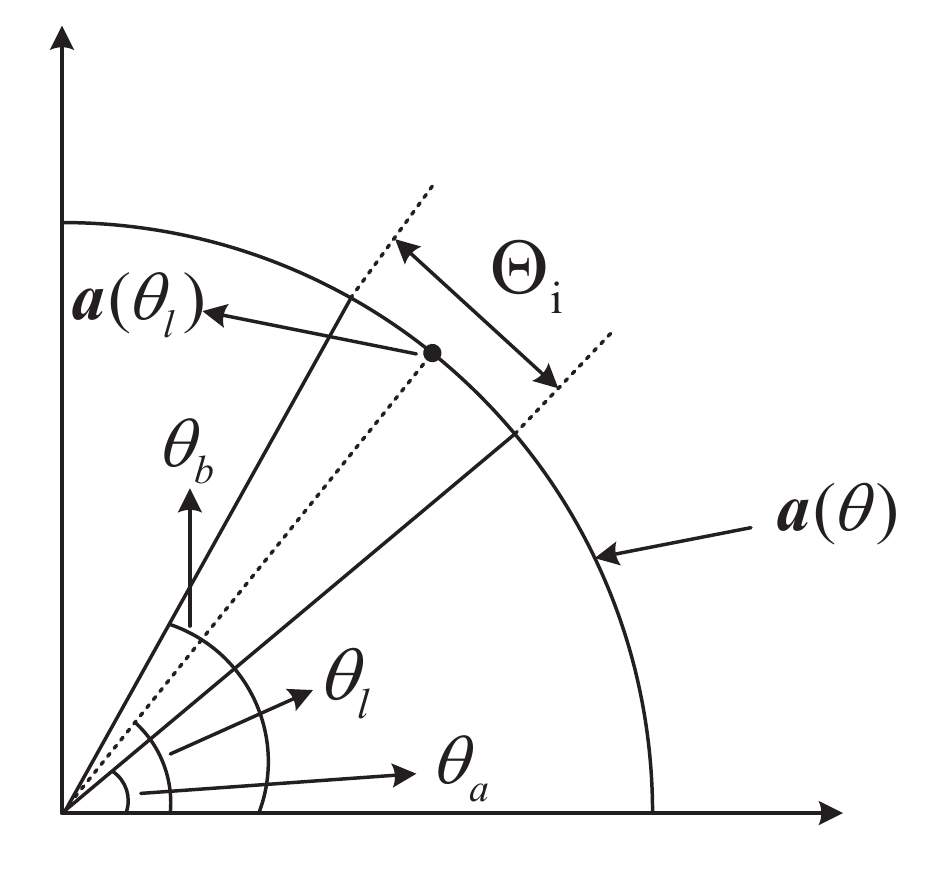}
		\caption{The concept of the direction range $\Theta_{\mathrm{i}}$ and the discretized angle $\theta_{i}$.}
		\label{angle}
	\end{figure}

	Second, as shown in Fig.~\ref{angle}, with the curve representing the steering vector $\boldsymbol{a}(\theta)$, the IPNCM is then reconstructed by integrating the Capon spatial spectrum over the range $\Theta_{\mathrm{i}}$ between $\theta_{a}$ and $\theta_{b}$ 
	\begin{equation}
		\label{eq11}
		\begin{aligned}
			\hat{\boldsymbol{R}}_{\mathrm{INF}}&=\oint_{\Theta_{\mathrm{i}} }{P(\theta)\boldsymbol{a}(\theta)\boldsymbol{a}^{\mathrm{H}}(\theta)d\theta}\\
			&=\oint_{\Theta_{\mathrm{i}} }{\frac{\boldsymbol{a}(\theta)\boldsymbol{a}^{\mathrm{H}}(\theta)}{\boldsymbol{a}^{\mathrm{H}}(\theta)\hat{\boldsymbol{R}}^{-1}\boldsymbol{a}(\theta)} d\theta},
		\end{aligned}
	\end{equation}
	where $\Theta_{\mathrm{i}}$ is the interference range, and the desired signal is not included. Considering that the accuracy of the direction of arrival (DOA) algorithm is affected by the resolution of the array and propagation environment, in this paper, we set the range of the  interference $\hat{\theta}_{\mathrm{i}}$ as $[\hat{\theta}_{\mathrm{i}}-8^{\circ}, \hat{\theta}_{\mathrm{i}}+8^{\circ}]$, where $\hat{\theta}_{i}$ is the presumed interference direction and estimated by the DOA estimation algorithm.  $\hat{\boldsymbol{R}}_{\mathrm{INF}}$ collects all interference and noise information without the effect of the desired signal. 
	
	However, the computational complexity of the integral operation in (\ref{eq11}) is high and cannot be solved efficiently. Third, the integral operation can be approximated by a summation as \cite{GU2014375}
	\begin{equation}
		\label{eq12}
		\hat{\boldsymbol{R}}_{\mathrm{INF}}\approx \sum_{l=1}^{L} {\frac{\boldsymbol{a}(\theta_{l})\boldsymbol{a}^{\mathrm{H}}(\theta_{l})}{\boldsymbol{a}^{\mathrm{H}}(\theta_{l})\hat{\boldsymbol{R}}^{-1}\boldsymbol{a}(\theta_{l})}d\theta},
	\end{equation}
	where the range $\Theta_{\mathrm{i}}$ is discretized into $L$ angles, the $l$-th angle is denoted as $\theta_{l}$, and $\boldsymbol{a}({\theta_{l}})$ is the corresponding steering vector. The approximation accuracy is dependent on the discretization number $L$, and the computational complexity of this method is $\mathcal{O}(M^{2}L)$. 
	
	However, in this method, the IPNCM $\boldsymbol{R}_{\mathrm{INF}}$ is estimated from $\hat{\boldsymbol{R}}$ as shown in the denominator of (\ref{eq12}), where the desired signal component is included. Moreover, a large number of discretized angles are needed to approximate the integration operation precisely.
	
	\section{The Proposed URGLQ Algorithm}\label{secprop}
\begin{figure}[ht]
		\centering 
		\includegraphics[width=0.5\textwidth]{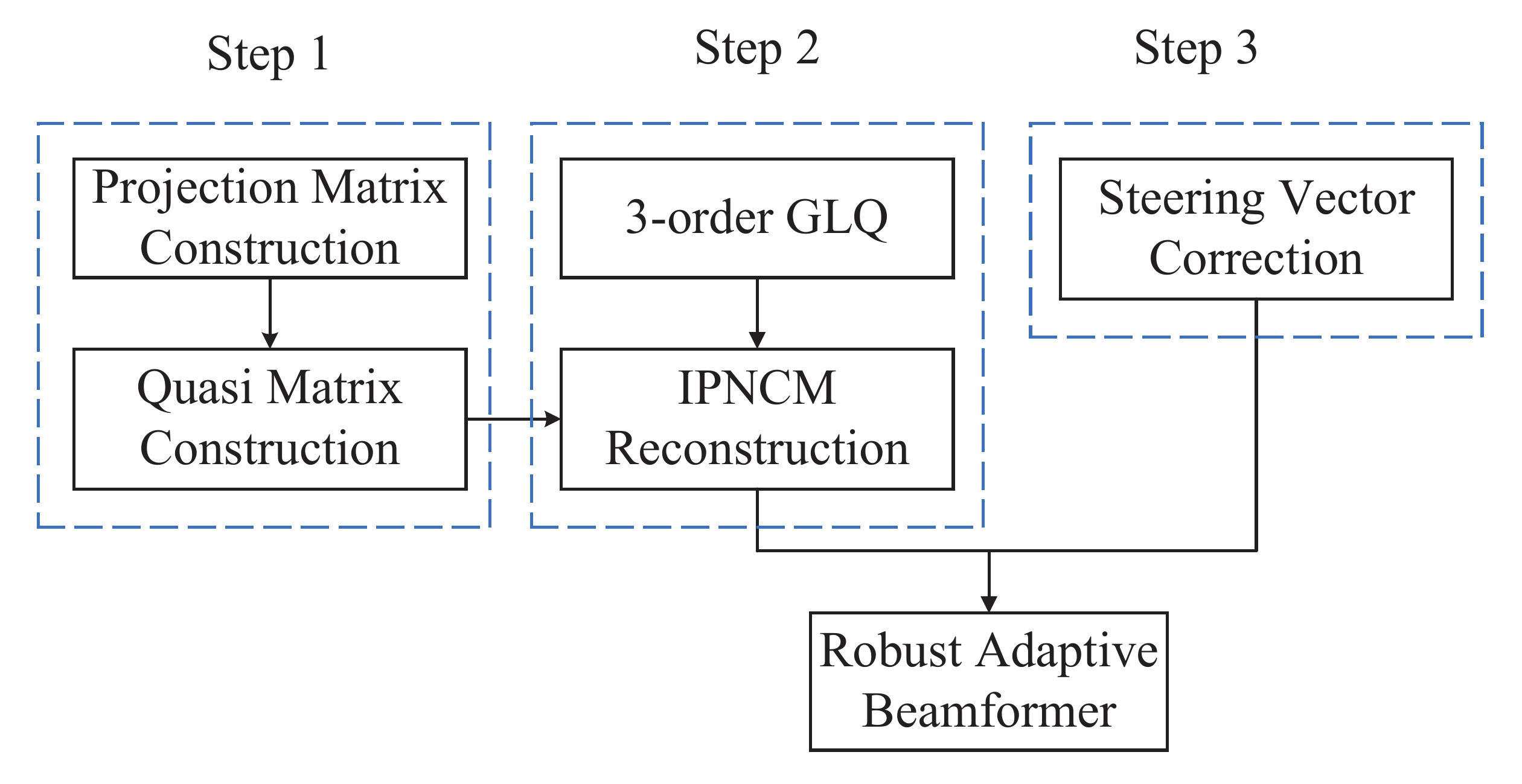}
		\caption{The proposed URGLQ algorithm flowchart.}
		\label{flow1}
	\end{figure}

	In this section, a novel adaptive beamforming algorithm based on IPNCM reconstruction is proposed, which contains $3$ steps as shown in Fig.~\ref{flow1}. First, a projection matrix is introduced to eliminate the desired signal from the received one. A quasi-covariance matrix with less desired information is obtained after the projection and replaces the SCM in (\ref{eq12}). Second, a matrix reconstruction method based on the Gauss-Legendre quadrature (GLQ) is proposed to replace the integral operation and reduce the computational complexity. At last, the steering vector is updated to reduce the model mismatch by maximizing the output power of the beamformer.
	
	\subsection{Step 1: Projection Matrix Construction}
	In this step, a projection matrix $\boldsymbol{B}$ is constructed to remove the desired signal from the received signal $\boldsymbol{x}(k)$, so we have $\boldsymbol{B}\boldsymbol{x}_{\text{s}}(k)\rightarrow \boldsymbol{0}$. Then, the IPNCM can be estimated more accurately. First, we construct a covariance-like matrix as
	\begin{equation}
		\label{eigen}
		\boldsymbol{C} = \alpha\boldsymbol{a}(\theta_{0})\boldsymbol{a}^{\mathrm{H}}(\theta_{0})+\boldsymbol{\mathrm{I}}_{M},
	\end{equation}
	where $\alpha\gg 1 $ is a construction parameter. To ensure good orthogonality between the constructed matrix and the desired signal, we set $\alpha=100\operatorname{Tr}\left \{ \hat{R} \right \}$ in simulations of this paper, and it can be set as $\alpha = 10000$ in the practical application to simplify the calculation. With the eigenvalue decomposition, we have
	\begin{equation}
		\boldsymbol{C} = \sum_{i=1}^{M}\mu_{i}\boldsymbol{p}_{i}\boldsymbol{p}_{i}^{\mathrm{H}},
	\end{equation}
	where $\mu_{i}$ $(i = 1,2,\dots,M)$ denotes the eigenvalue in descending order, and the eigenvalues can be obtained as $\mu_{1}=M\alpha+1$ and $\mu_{i} = 1$. $\boldsymbol{p}_{i}$ is the eigenvector corresponding to $\mu_{i}$. Then, the inverse matrix can be expressed as
	\begin{equation}
		\boldsymbol{C}^{-1} 
		= \sum_{i=1}^{M} \frac{\boldsymbol{p}_{i}\boldsymbol{p}_{i}^{\mathrm{H}}}{\mu_{i}}
		= \frac{\boldsymbol{p}_{1}\boldsymbol{p}_{1}^{\mathrm{H}}}{M\alpha +1} + \sum_{i=2}^{M}\boldsymbol{p}_{2}\boldsymbol{p}_{2}^{\mathrm{H}}.
	\end{equation}
	Considering that $\sum_{i=1}^{M}\boldsymbol{p}_{i}\boldsymbol{p}_{i}^{\mathrm{H}}=\boldsymbol{\mathrm{I}}_{M}$ is fulfilled and $\alpha \gg 1$, $\boldsymbol{C}^{-1}$ can be further approximated as
	\begin{equation}
		\boldsymbol{C}^{-1} 
		\approx  \sum_{i=2}^{M}\boldsymbol{p}_{i}\boldsymbol{p}_{i}^{\mathrm{H}} = \boldsymbol{\mathrm{I}}_{M}-\boldsymbol{p}_{1}\boldsymbol{p}_{1}^{\mathrm{H}}.
	\end{equation}
	Hence, the projection matrix $\boldsymbol{B}$ can be chosen as 
	\begin{equation}
		\label{B}
		\boldsymbol{B} = \boldsymbol{\mathrm{I}}_{M}-\boldsymbol{p}_{1}\boldsymbol{p}_{1}^{\mathrm{H}}.
	\end{equation}
	
\begin{figure}
		\centering 
		\includegraphics[width=0.4\textwidth]{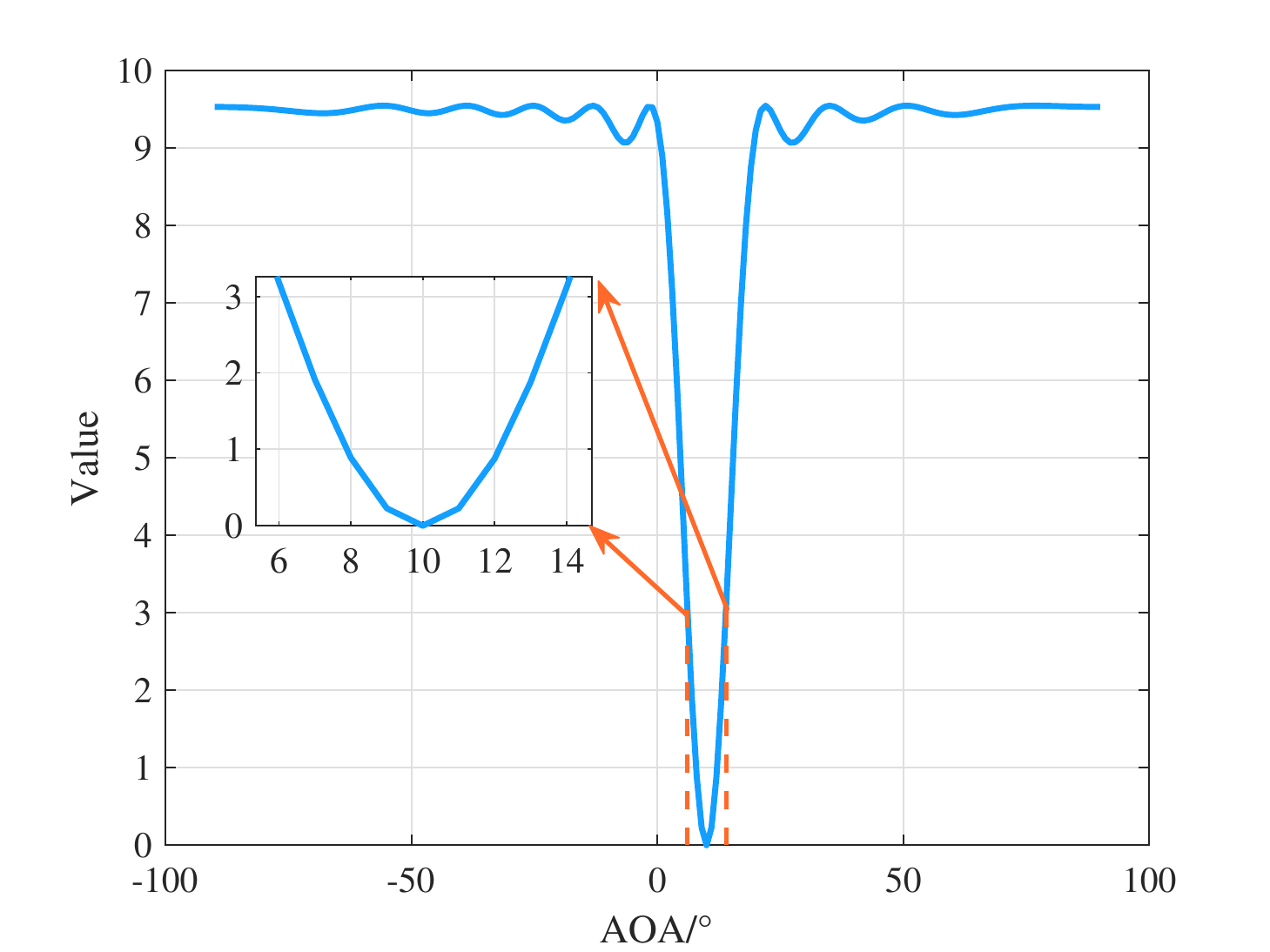}
		\caption{The value of  $\parallel \boldsymbol{B}^{\mathrm{H}}\boldsymbol{a}(\theta)\parallel_{2}^{2}$ versus $\theta$, where  $\alpha = 100\operatorname{Tr}\left\{\hat{\boldsymbol{R}}\right\}$}
		\label{function}
	\end{figure}

	A function $\|\boldsymbol{B}^{\mathrm{H}}\boldsymbol{a}(\theta)\|^2_2$ versus $\theta$ is plotted in Fig.~\ref{function} to show that the desired signal can be removed efficiently with the projection matrix $\boldsymbol{B}$. In this example, 10 sensors are considered. $1$ desired signal comes from the direction $10^{\circ}$ and $2$ interference is assumed to be $-30^{\circ}$ and $40^{\circ}$. The signal-to-noise ratio (SNR) and interference-to-noise ratio (INR) in each sensor are both set as $20$~dB, the parameter $\alpha = 10^{2}\operatorname{Tr}\{\hat{\boldsymbol{R}}\}$. As shown in Fig.~\ref{function}, $\|\boldsymbol{B}^{\mathrm{H}}\boldsymbol{a}(\theta)\|^2_2$ is small when $\theta$ is located in the direction of the desired signal, which means that the desired signal is removed effectively.
	
	Using the projection matrix $\boldsymbol{B}$, the quasi-matrix $\tilde{\boldsymbol{R}}$ containing almost only the interference and noise information can be calculated as
	\begin{equation}
		\begin{aligned}
			\tilde{\boldsymbol{R}} &= \frac{1}{K} \sum_{k=1}^{K}\boldsymbol{B}^{\mathrm{H}}\boldsymbol{x}(k)\left[\boldsymbol{B}^{\mathrm{H}}\boldsymbol{x}(k)\right]^{\text{H}} + \hat{\sigma}_{\mathrm{n}}^{2}\boldsymbol{\mathrm{I}}_{M},\\
			&= \boldsymbol{B}^{\mathrm{H}}\hat{\boldsymbol{R}}\boldsymbol{B} + \hat{\sigma}_{\mathrm{n}}^{2}\boldsymbol{\mathrm{I}}_{M},
		\end{aligned}
	\end{equation}  
	where $\hat{\sigma}_{\mathrm{n}}^{2}$ is the estimate of the noise power, and is usually computed as the minimum eigenvalue of SCM. Then, the IPNCM can be reconstructed by replacing the $\hat{\boldsymbol{R}}$ in (\ref{eq11}) with $	\tilde{\boldsymbol{R}}$
	\begin{equation}
		\label{ipncm2}
		\hat{\boldsymbol{R}}_{\mathrm{INF}}
		=\oint_{\Theta_{\mathrm{i}} }{\frac{\boldsymbol{a}(\theta)\boldsymbol{a}^{\mathrm{H}}(\theta)}{\boldsymbol{a}^{\mathrm{H}}(\theta) \tilde{\boldsymbol{R}}^{-1}\boldsymbol{a}(\theta)} }d\theta.
	\end{equation}
	
	\subsection{Step 2: IPNCM Reconstruction Using Gauss-Legendre Quadrature}
	Since the complexity of (\ref{eq12}) grows rapidly as the number of discretization $L$ increases, an efficient matrix reconstruction method based on the Gauss-Legendre quadrature is proposed. 
	
	Gauss-Legendre quadrature is a kind of Gauss interpolation integral formula and is one of the highest algebraic precision methods in the interpolation-type quadrature formulas~\cite{laurie2001computation,prentice2006comments}.  Additionally, for the integrands which can be well approximated by polynomials, the Gauss-Legendre quadrature can approximate operation with high precision~\cite{johansson2018fast}. A generalized interpolation-type Gaussian quadrature formula with $N$ points can be written as
	\begin{equation}
		\label{Gau}
		\int_{a}^{b}f(z)dz = \sum_{n=0}^{N-1}A_{n}f(z_{n}) + E,
	\end{equation}
	where $z_{n}\in[a,b]$ $(n=0,1,\dots,N-1)$ is the Gauss integral node, $a\in\mathbb{R}$ and $b\in\mathbb{R}$ are the lower and upper bounds of the integral respectively, $E$ is the integral remainder (residual error), and $A_{n}$ is the weight, which can be calculated as 
	\begin{equation}
		\label{xishu}
		A_{n} = \int_{a}^{b}\frac{h(z)}{(z-z_{n})\frac{\partial h(z)}{\partial z} }dz
	\end{equation}
	where $h(z)=(z-z_{0})(z-z_{1})\cdots(z-z_{n})$ is a polynomial in $z$. It is worth mentioning that the Gauss theorem states that $h(z)$ must be orthogonal to any polynomials of less than power $N$ \cite{S1064827500379690}.
	
	Moreover, the $N$-th normal Legendre polynomial is written as
	\begin{equation}
		\label{leg}
		P_{N}(z) = \frac{1}{2^{N}N!}\frac{\partial ^{N} (z^{2}-1)^{N}}{\partial  z^{N}}.
	\end{equation}
	We can choose the roots of the Legendre polynomial as the integral nodes in the Gaussian quadrature formula (\ref{Gau}), i.e., letting $z_n$ as the roots of $P_{N}(z)=0$, and construct the GLQ.
	
	Taking the trade-off between computational complexity and algebraic precision into account, we can choose $3$-order GLQ~\cite{shuai2021fast} to calculate the integral. Additionally, the integral interval is chosen as $[-1, 1]$ to facilitate calculation, i.e., $a=-1$ and $b=1$. Then, Eq.~($\ref{Gau}$) can be rewritten as
	\begin{equation}\label{eq23}
		\int_{-1}^{1}f(z)dz = A_{0}f(z_{0})+ A_{1}f(z_{1}) + A_{2}f(z_{2}) + E,
	\end{equation}
	where $z_{0} = -\sqrt{15}/5$, $z_{1} = 0$, $z_{2} = \sqrt{15}/5$ are the roots of the $3$-order Legendre polynomial, i.e.,
	\begin{equation}  
		P_{3}(z) = \frac{1}{2}(5z^{3}-3z)=0. 
	\end{equation}
	
	Next, letting $h(z) = P_{N}(z)$, the weights $A_n$ in (\ref{xishu}) can be further obtained as 
	\begin{equation}
		\begin{aligned}
			A_{n} &= \int_{-1}^{1}\frac{P_{N}(z)}{(z-z_{n})\frac{\partial P_{N}(z_{n)}}{\partial  z}} dz = \frac{2}{(1-z_{n}^{2})\left[\frac{\partial P_{N}(z_{n)}}{\partial  z}\right]^{2}}.
		\end{aligned}
	\end{equation}
	Thus, using $3$-order  Legendre polynomial, the weights can be obtained as $A_{0} = 5/9$, $A_{1} = 8/9$, $A_{2} = 5/9$. 
	
	Finally, by combining the roots $x_n$ and the weights $A_n$, Eq.~(\ref{eq23}) can be approximated as
	\begin{equation}
		\label{3order}
		\int_{-1}^{1}f(z)dz \approx \frac{5}{9}f\left(\frac{-\sqrt{15}}{5}\right) + \frac{8}{9}f(0) + \frac{5}{9}f\left(\frac{\sqrt{15}}{5}\right) 
	\end{equation}
	Furthermore, using a simple linear transformation, a general integral can be approximated by a $3$-order GLQ as
	\begin{equation}
		\int_{a}^{b}f(z)dz \approx \frac{b-a}{2}\left[\frac{5}{9}f(l_{0})+\frac{8}{9}f(l_{1})+\frac{5}{9}f(l_{2})\right],
	\end{equation}
	where $l_{n} = \frac{1}{2}(a+b)+\frac{1}{2}z_{n}(b-a)$.
	
	Therefore, the IPNCM (\ref{ipncm2}) can be calculated by the $3$-order GLQ as
	\begin{equation}
		\begin{aligned}
			\label{ipncm3}
			\hat{\boldsymbol{R}}_{\mathrm{INF}}
			&=\oint_{\Theta_{\mathrm{i}} }{\frac{\boldsymbol{a}(\theta)\boldsymbol{a}^{\mathrm{H}}(\theta)}{\boldsymbol{a}(\theta)^{H} \tilde{\boldsymbol{R}}_{\mathrm{INF}}^{-1}\boldsymbol{a}(\theta)} d\theta}\\
			&=\frac{\theta_{b}-\theta_{a}}{2}\left[\frac{5}{9}f(l_{0})+\frac{8}{9}f(l_{1})+\frac{5}{9}f(l_{2})\right],\\
			f(\theta) &= \frac{\boldsymbol{a}(\theta)\boldsymbol{a}^{\mathrm{H}}(\theta)}{\boldsymbol{a}^{\mathrm{H}}(\theta)\tilde{\boldsymbol{R}}_{\mathrm{INF}}^{-1}\boldsymbol{a}(\theta)},
		\end{aligned}
	\end{equation}
	where $\theta_{a}$ and $\theta_{b}$ are the lower and upper bounds of the integral angular range $\Theta_{\mathrm{i}}=[\theta_{a}, \theta_{b}]$, and $l_{n}=\frac{1}{2}(\theta_{a}+\theta_{b})+\frac{1}{2}x_{n}(\theta_{b}-\theta_{a})$. Additionally, when calculating the IPNCM by the 3-order GLQ, the integral angular range $\Theta_{\mathrm{i}}$ is the range of interference.

	\subsection{Step 3: Desired Signal Steering Vector Estimation}
	In practice, the presumed desired steering vector $\boldsymbol{a}(\theta_{0})$, which is simply obtained by the DOA estimation, is usually different from the actually desired steering vector $\hat{\boldsymbol{a}}(\theta_{0})$ due to the complex propagation environment. Therefore, a steering vector estimation method based on maximizing the beamformer output power is proposed in this step.
	
	To maximize the output SINR Eq.~(\protect\ref{sinr}), an optimization problem in Eq.~(\protect\ref{beamforming probelm}) can be obtained and solved by Eq.~(\protect\ref{mvdr}).  By substituting both the Capon beamformer (\ref{mvdr}) and the reconstructed IPNCM $\hat{\boldsymbol{R}}_{\mathrm{INF}}$ (\ref{ipncm3}) into Eq.~(\ref{beamforming probelm}), the output power of the beamformer can be obtained as \cite{gu2012robust}
	\begin{equation}
		P(\theta_{0}) = \frac{1}{\boldsymbol{a}(\theta_{0})^{\mathrm{H}}\hat{\boldsymbol{R}}^{-1}_{\mathrm{INF}}\boldsymbol{a}(\theta_{0})},
	\end{equation}
	which is a function of the steering vector $\boldsymbol{a}(\theta_{0})$, and $\boldsymbol{a}(\theta_{0})$ can be estimated by maximizing $P(\theta_{0})$. Suppose that $\hat{\boldsymbol{a}}(\theta_{0}) = \boldsymbol{a}(\theta_{0}) + \boldsymbol{e}$, where $\boldsymbol{e}$ is the mismatch vector, and the optimization problem of estimating $\hat{\boldsymbol{a}}(\theta_{0})$ can be expressed as
	\begin{align}
		\label{opt1}
		&\underset{\boldsymbol{e}}{\min}&&(\boldsymbol{a}(\theta_{0}) + \boldsymbol{e})^{\mathrm{H}}\hat{\boldsymbol{R}}_{\mathrm{INF}}^{-1}(\boldsymbol{a}(\theta_{0}) + \boldsymbol{e})\\
		&\mathrm{s.t.}&&(\boldsymbol{a}(\theta_{0}) + \boldsymbol{e})^{\mathrm{H}}\hat{\boldsymbol{R}}_{\mathrm{INF}}(\boldsymbol{a}(\theta_{0}) + \boldsymbol{e}) \le \boldsymbol{a} ^{\mathrm{H}}(\theta_{0})\hat{\boldsymbol{R}}_{\mathrm{INF}}\boldsymbol{a}(\theta_{0}),\nonumber
	\end{align}
	where the constraint prevents the estimated steering vector $\boldsymbol{a}$ from converging to the range $\Theta_{\mathrm{i}}$.
	
	The mismatch vector $\boldsymbol{e}$ can be further decomposed into $\boldsymbol{e}_{\perp }$ and $\boldsymbol{e}_{\parallel }$, where $\boldsymbol{e}_{\perp }$ is orthogonal to $\boldsymbol{a}(\theta_{0})$, while the $\boldsymbol{e}_{\parallel }$ is parallel to $\boldsymbol{a}(\theta_{0})$. Consider that $\boldsymbol{e}_{\parallel }$ is a scaled copy of $\hat{\boldsymbol{a}}(\theta_{0})$, so it does not affect the beamforming performance. Thus, (\ref{opt1}) can be transformed into
	\begin{align}
		\label{opt2}
		&\underset{\boldsymbol{e}_{\perp}}{\min}&&(\boldsymbol{a}(\theta_{0}) + \boldsymbol{e}_{\perp})^{\mathrm{H}}\hat{\boldsymbol{R}}_{\mathrm{INF}}^{-1}(\boldsymbol{a}(\theta_{0}) + \boldsymbol{e}_{\perp}) \nonumber\\
		&\mathrm{s.t.}&&\boldsymbol{a}^{\mathrm{H}}(\theta_{0})\boldsymbol{e}_{\perp} = 0\\
		& &&(\boldsymbol{a}(\theta_{0}) + \boldsymbol{e}_{\perp})^{\mathrm{H}}\hat{\boldsymbol{R}}_{\mathrm{INF}}(\boldsymbol{a}(\theta_{0}) + \boldsymbol{e}_{\perp}) \le \boldsymbol{a} ^{\mathrm{H}}(\theta_{0})\hat{\boldsymbol{R}}_{\mathrm{INF}}\boldsymbol{a}(\theta_{0}),\nonumber
	\end{align}
	where the equality constraint maintains the orthogonality between $\boldsymbol{e}_{\perp}$ and $\overline{\boldsymbol{a}}$. Since the optimization problem (\ref{opt2}) is a feasible quadratically constrained quadratic programming (QCQP) problem, it can be efficiently solved by a convex optimization toolbox.
	
	Finally, substituting the reconstructed IPNCM $\hat{\boldsymbol{R}}_{\mathrm{INF}}$ with the estimated steering vector $\hat{\boldsymbol{a}}(\theta_{0})$ back into (\ref{mvdr}), the adaptive beamformer can be calculated as
	\begin{equation}
		\boldsymbol{w}_{\mathrm{prop}} = \frac{\hat{\boldsymbol{R}}_{\mathrm{INF}}^{-1}\hat{\boldsymbol{a}}(\theta_{0})}{\hat{\boldsymbol{a}}^{\mathrm{H}}(\theta_{0})\hat{\boldsymbol{R}}_{\mathrm{INF}}^{-1}\hat{\boldsymbol{a}}(\theta_{0})}.
	\end{equation}
	
	Based on the above discussion, the detailed procedures of the proposed adaptive beamforming algorithm are described in Algorithm~\ref{alg1}. In the proposed URGLQ algorithm, matrix $\tilde{\boldsymbol{R}}$ is obtained from received data by projecting SCM $\hat{\boldsymbol{R}}$ onto $\boldsymbol{C}$. Then, an IPNCM reconstruction method based GLQ and $\tilde{\boldsymbol{R}}$ is proposed to reduce the computational complexity and maintain excellent performance. Combining the reconstructed IPNCM $\hat{\boldsymbol{R}}$ and corrected desired steering vector $\hat{\boldsymbol{a}}(\theta_{0})$, the proposed beamformer is obtained.
	
	\begin{algorithm}[t]
		\caption{Proposed URGLQ Adaptive Beamforming} 
		\label{alg1}
		\begin{algorithmic}[1]
			\STATE \textbf{Input:} Array received data $\left\{ \boldsymbol{x}(k)\right\}_{k=1}^{K}$\\
			\STATE Calculate the SMI $\hat{\boldsymbol{R}}=\frac{1}{K}\sum_{k=1}^{K}\boldsymbol{x}(k)\boldsymbol{x}(k)_{\mathrm{H}}$;
			\STATE Construct the covariance-like matrix $\boldsymbol{C}=\alpha\boldsymbol{a}(\theta_{0})\boldsymbol{a}^{\mathrm{H}}(\theta_{0})+\boldsymbol{\mathrm{I}}_{M}$, $\alpha = 100\operatorname{Tr}\left\{\hat{\boldsymbol{R}}\right\}$;
			\STATE Construct the projection matrix $\boldsymbol{B} =\boldsymbol{C}^{-1}= \boldsymbol{\mathrm{I}}_{M} \boldsymbol{p}_{1}\boldsymbol{p}_{1}^{\mathrm{H}}$;
			\STATE Calculate quasi-matrix $\tilde{\boldsymbol{R}} = \boldsymbol{B}^{\mathrm{H}}\hat{\boldsymbol{R}}\boldsymbol{B} + \hat{\sigma}_{\mathrm{n}}^{2}\boldsymbol{\mathrm{I}}_{M}$;\\
			\STATE Reconstruction the IPNCM $\hat{\boldsymbol{R}}_{\mathrm{INF}}$ based on GLQ, $\hat{\boldsymbol{R}}_{\mathrm{INF}}=\frac{\theta_{b}-\theta_{a}}{2}[A_{0}f(l_{0})+A_{1}f(l_{1})+A_{2}f(l_{2})]$,
			$f(\theta) = \frac{\boldsymbol{a}(\theta)\boldsymbol{a}^{\mathrm{H}}(\theta)}{\boldsymbol{a}^{\mathrm{H}}(\theta)\tilde{\boldsymbol{R}}^{-1}\boldsymbol{a}(\theta)}, \theta \in \Theta_{\mathrm{i}}$;
			\STATE Correct the desired signal steering vector by solving a QCQP (\ref{opt2});
			\STATE Design proposed adaptive beamformer $\boldsymbol{w}_{\mathrm{prop}} = \frac{\hat{\boldsymbol{R}}_{\mathrm{INF}}\hat{\boldsymbol{a}}(\theta_{0})}{\hat{\boldsymbol{a}}^{\mathrm{H}}(\theta_{0})\hat{\boldsymbol{R}}_{\mathrm{INF}}\hat{\boldsymbol{a}}(\theta_{0})}$;
			\STATE \textbf{Output:} Proposed robust adaptive beamforming weight vector $\boldsymbol{w}_{\mathrm{prop}}$
		\end{algorithmic}
	\end{algorithm}
	
	\subsection{Computational Complexity Analysis}
	Since the proposed beamforming algorithm consists of three parts: projection matrix construction, IPNCM reconstruction using Gauss-Legendre quadrature and desired signal steering vector estimation, its computational complexity is determined by the three parts. Furthermore, the IPNCM reconstruction part, which is based on the Gauss-Legendre quadrature,  only needs three addition operations and has relatively low complexity. Thus, the computational complexity of the proposed algorithm mainly depends on the projection matrix construction and steering vector estimation. In projection matrix construction, the computational complexity is determined by the eigenvalue decomposition operation of the covariance-like matrix $\boldsymbol{C}$ (\ref{eigen}), which is $\mathcal{O}(M^{3})$. In steering vector estimation, a QCQP (\ref{opt2}) is solved, and the complexity is $\mathcal{O}(M^{3.5})$. Thus, the computational complexity of the proposed algorithm is $\mathcal{O}(M^{3.5})$. Meanwhile, to reconstruct the IPNCM, the computational complexity of LINEAR~\cite{gu2012robust}, VOLUME~\cite{huang2015robust}, and ISVPE~\cite{zheng2018covariance} is $\mathcal{O}(\mathrm{max}(LM^{2},M^{3.5}))$, while SUB~\cite{yuan2017robust} is $\mathcal{O}(\mathrm{max}(LM^{2},M^{3}))$ and MEPS~\cite{zheng2019robust} is $\mathcal{O}(LM^{2})$. Note that, the $L$ denotes the number of discretization points in the angular domain, which is much larger than the number of sensors $M$ in general. Thus, the complexity of the proposed algorithm is much less than most beamformers.
	
	\section{Simulation and Experimental Results}\label{secsim}
	
	The parameters in this simulation are shown in Table~\ref{table1}. In this part, the ULA with $M=10$ omnidirectional sensors, which are spaced half-wavelength, is considered. For each antenna sensor, the additive noise is modeled as a complex Gaussian process, which is spatially and temporally white with zero mean and unit variance. One desired signal and two interference are impinging from $\theta_{0}=10^{\circ}$, $\theta_{1}=-30^{\circ}$ and $\theta_{2}=40^{\circ}$, respectively. In all the examples, the INR is set as $20$ dB. The interference and desired signal ranges are set as $\Theta_{\mathrm{i}}=[\theta_{1}-8^{\circ}, \theta_{1}+8^{\circ}]\cup\theta_{2}-8^{\circ},\theta_{2}+8^{\circ}]$ and $\Theta_{\mathrm{s}} = [\theta_{0}-8^{\circ}, \theta_{0}+8^{\circ}]$. The number of discretizations $L$ is set as $20$ in all examples except example 1. The simulation parameters are summarized in TABLE \ref{table1}. The code is available online \url{https://github.com/chenpengseu/robust-adaptive-beamforming-2022.git}.

	\begin{table}[!ht]
		\centering
		\caption{The parameters of the simulation}
		\begin{tabular}{c c}
			\toprule     
			Parameter & Value  \\ 
			\midrule   
			The number of sensors $M$ & 10 \\
			The spacing $d$ & half-wavelength  \\
			DOA of desired signal $\theta_{0}$ & $10^{\circ}$  \\
			DOA of interference $(\theta_{1}$, $\theta_{2})$ & $(-30^{\circ}$, $40^{\circ})$\\
			INR & $20$ dB\\
			Desired signal range $\Theta_{\mathrm{s}}$ & $\left[ \theta_{0}-8^{\circ}, \theta_{0}+8^{\circ}\right]$\\
			Interference range $\Theta_{\mathrm{s}}$ & $\left[ \theta_{1}-8^{\circ}, \theta_{1}+8^{\circ}\right]\cup\theta_{2}-8^{\circ},\theta_{2}+8^{\circ}]$\\
			The number of discretizations $L$ & $20$\\
			\bottomrule   
		\end{tabular}
		\label{table1}
	\end{table}
	
	The proposed algorithm URGLQ is compared with the following $5$ different matrix reconstruction methods: 
	\begin{itemize}
		\item The optimal algorithm MVDR~\cite{capon1969high}, which assumes the desired signal steering vector and antenna array structure are known precisely; 
		\item Interference covariance matrix reconstruction beamformer (LINEAR) \cite{gu2012robust}, which is the first one to reconstruct the IPNCM by linear integration over the Capon spectrum, where the parameter is $\epsilon =\sqrt{0.1}$ in LINEAR;
		\item  The covariance matrix reconstruction beamformer based on volume integration (VOLUME)~\cite{huang2015robust}, which replaces the linear integration with a volume integration based on LINEAR, where the parameters are $\varepsilon=0.3$, $\phi _{m}^{l}\subseteq [0,\pi]$, and $\rho=0.9$; 
		\item The interference covariance matrix reconstruction beamformer based on annular uncertainty set (AUS)~\cite{yang2020robust}, which estimates the desired steering vector by vector space projection, and improves the robustness of the algorithm, where $\rho=0.9$, $\phi _{m}^{l}\subseteq [0,\pi]$; 
		\item The covariance matrix reconstruction based on subspace projection (SUB)~\cite{yuan2017robust}, which reconstructs the IPNCM based on subspace and eigenvalue decomposition; 
		\item The covariance matrix reconstruction with maximum entropy spectrum (MEPS)~\cite{mohammadzadeh2020maximum}, which replaces the Capon spectrum in~\cite{gu2012robust} with maximum entropy spectrum to improve the performance of beamformer.
	\end{itemize}  
	
	All the optimization problems in the simulation and experiment are solved by the convex optimization toolbox CXV~\cite{grant2009cvx}. $300$ Monte Carlo simulations are performed in each scenario and the desired signal is always included in the received data of each snapshot. The number of snapshots is fixed to $K=30$ when the SNR changes. Similarly, the SNR is fixed to $20$ dB when the number of snapshots changes.
	
	\subsection{Example 1: The Comparison Between the Polynomial Summation and GLQ.}
	First, to verify the effectiveness of GLQ algorithm, we compare the performance of GLQ with the polynomial summation algorithm versus the number of discretizations. In this example, the input SNR and INR are set as $20$ dB. 
	\begin{figure}[h]
		\centering 
		\includegraphics[width=0.5\textwidth]{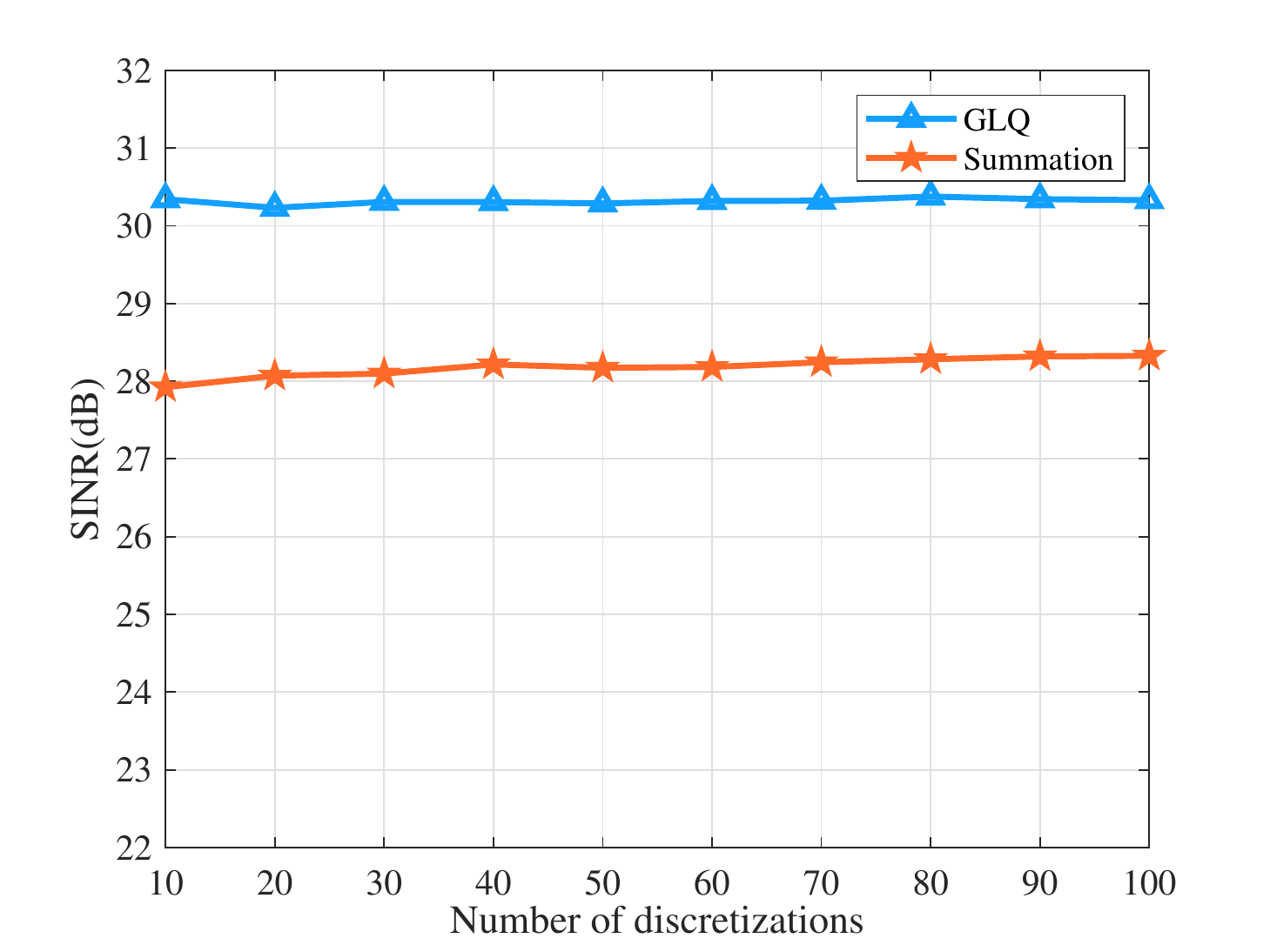}
		\caption{The performance comparison of GLQ method and summation versus the number of discretizations.}
		\label{COM}
	\end{figure}
	
	The performance comparison between the proposed method and polynomial summation is plotted in Fig.~\ref{COM}. As it shows, compared with the GLQ, the performance of the polynomial summation method is consistently worse, while the complexity is much higher than GLQ. Furthermore, with the increase in the number of discretizations, the performance of the polynomial summation method improves slowly, which means a sufficiently large number of discretizations is needed to achieve the same performance as GLQ. Based on this, the proposed algorithm has better precision with an approximate integral operation, while the integrands and power spectral density of the incident signal is not smooth. Thus, compared with the normal polynomial summation method with the computational complexity $\mathcal{O}(M^{2}L)$, the proposed GLQ method only needs three addition operations but gets much more excellent performance. By using GLQ, the computational complexity of the proposed adaptive beamforming is greatly reduced while the algorithm maintains high performance.
	
	\subsection{Example 2: Random Signal and Interference DOA Mismatch}
	First, the effect of random DOA mismatch is considered. The mismatch of both the desired signal and interference is uniformly distributed in [$-4^{\circ}$, $4^{\circ}$]. Thus, the desired signal is $\theta_{0}\in[6^{\circ}, 14^{\circ}]$, two interference are $\theta_{1}\in[-34^{\circ}, -26^{\circ}]$ and $\theta_{2}\in[36^{\circ}, 44^{\circ}]$, respectively. The DOA changes from run to run but keeps fixed from snapshot to snapshot. 
	\begin{figure}[t]
		\centering 
		\includegraphics[width=0.5\textwidth]{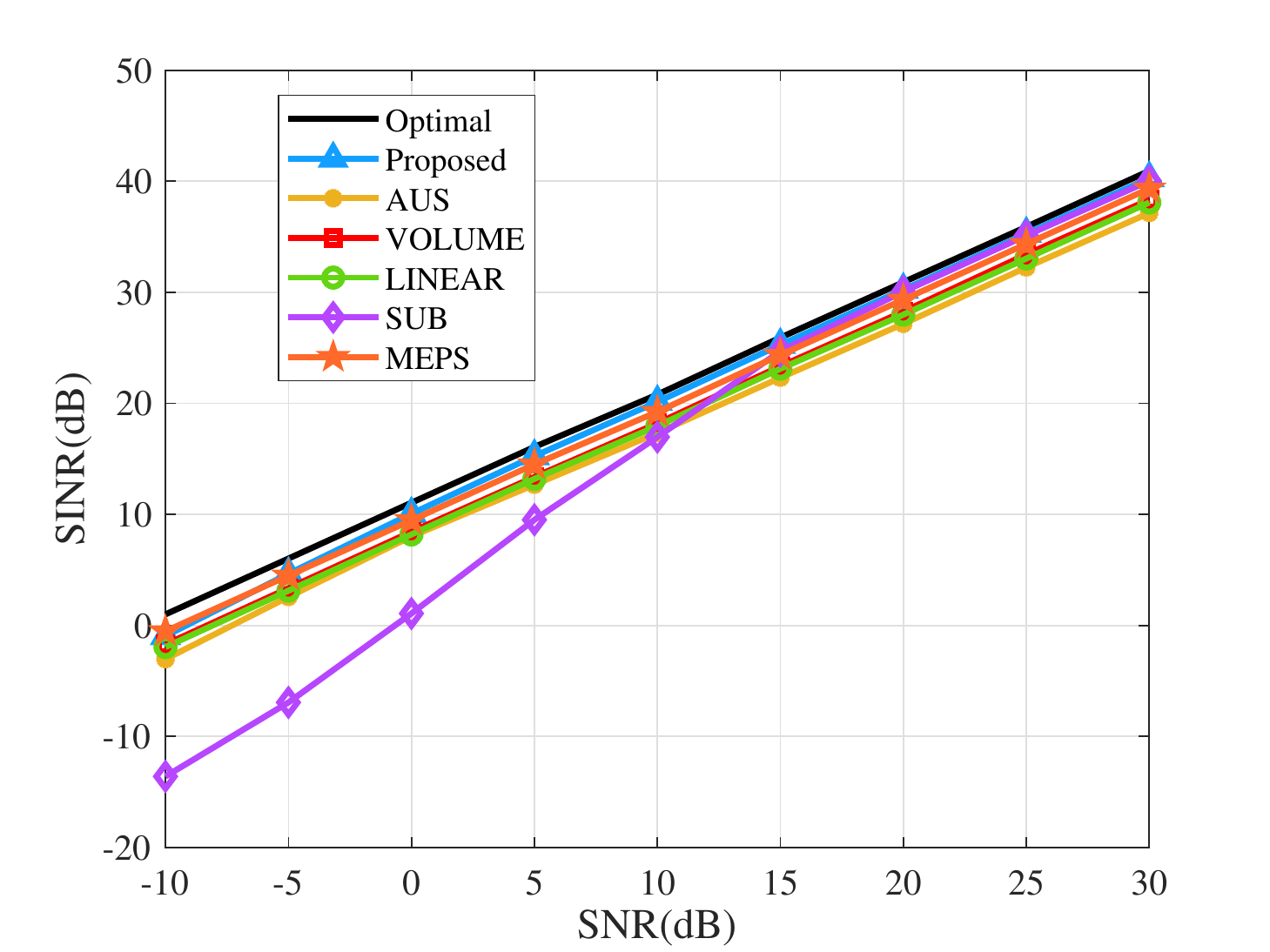}
		\caption{Output SINR versus input SNR in example 1.}
		\label{ep11}
	\end{figure}
	\begin{figure}[t]
		\centering 
		\includegraphics[width=0.5\textwidth]{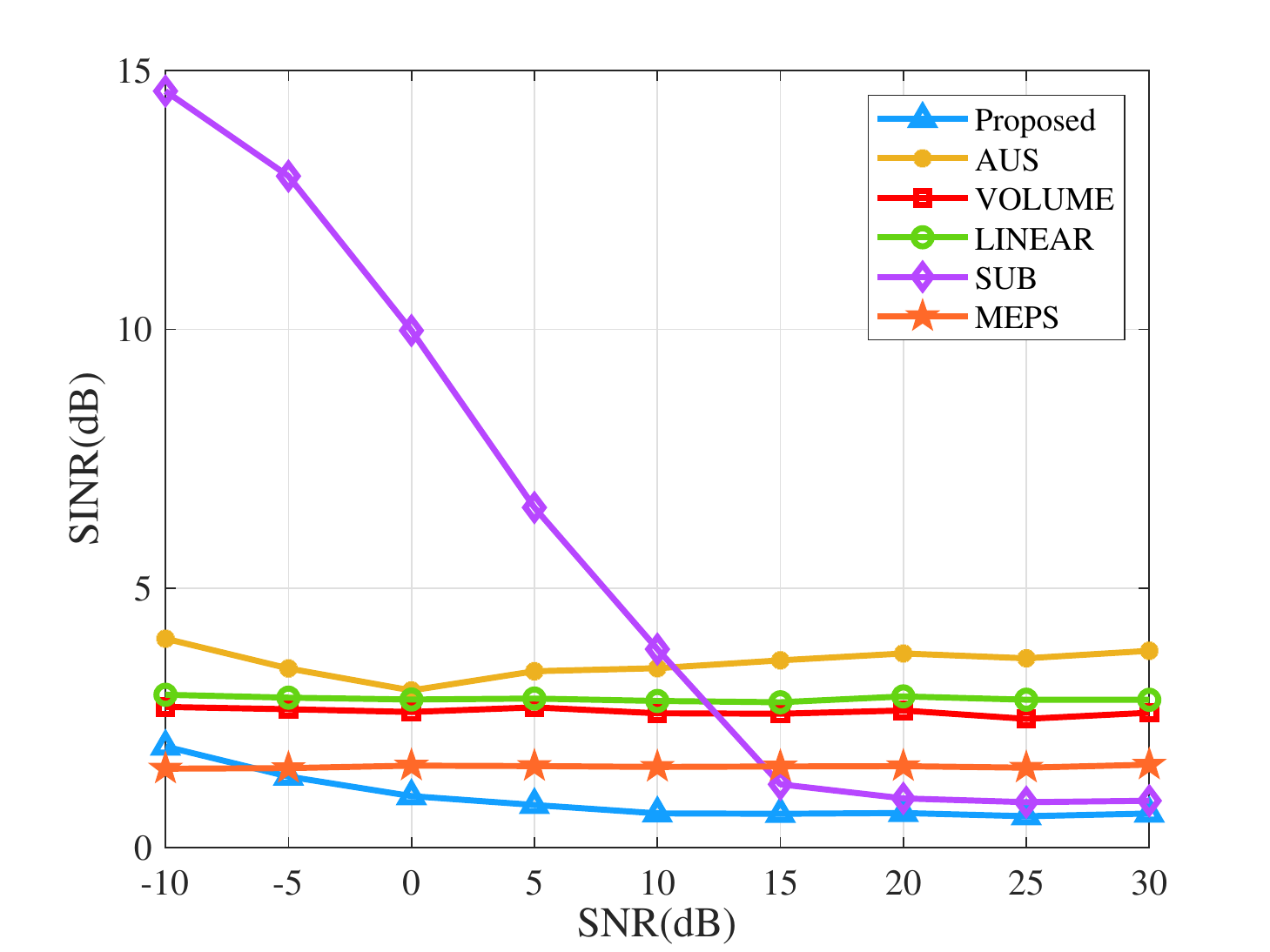}
		\caption{Deviation from optimal SINR versus SNR in example 1.}
		\label{ep12}
	\end{figure}
	\begin{figure}
		\centering 
		\includegraphics[width=0.5\textwidth]{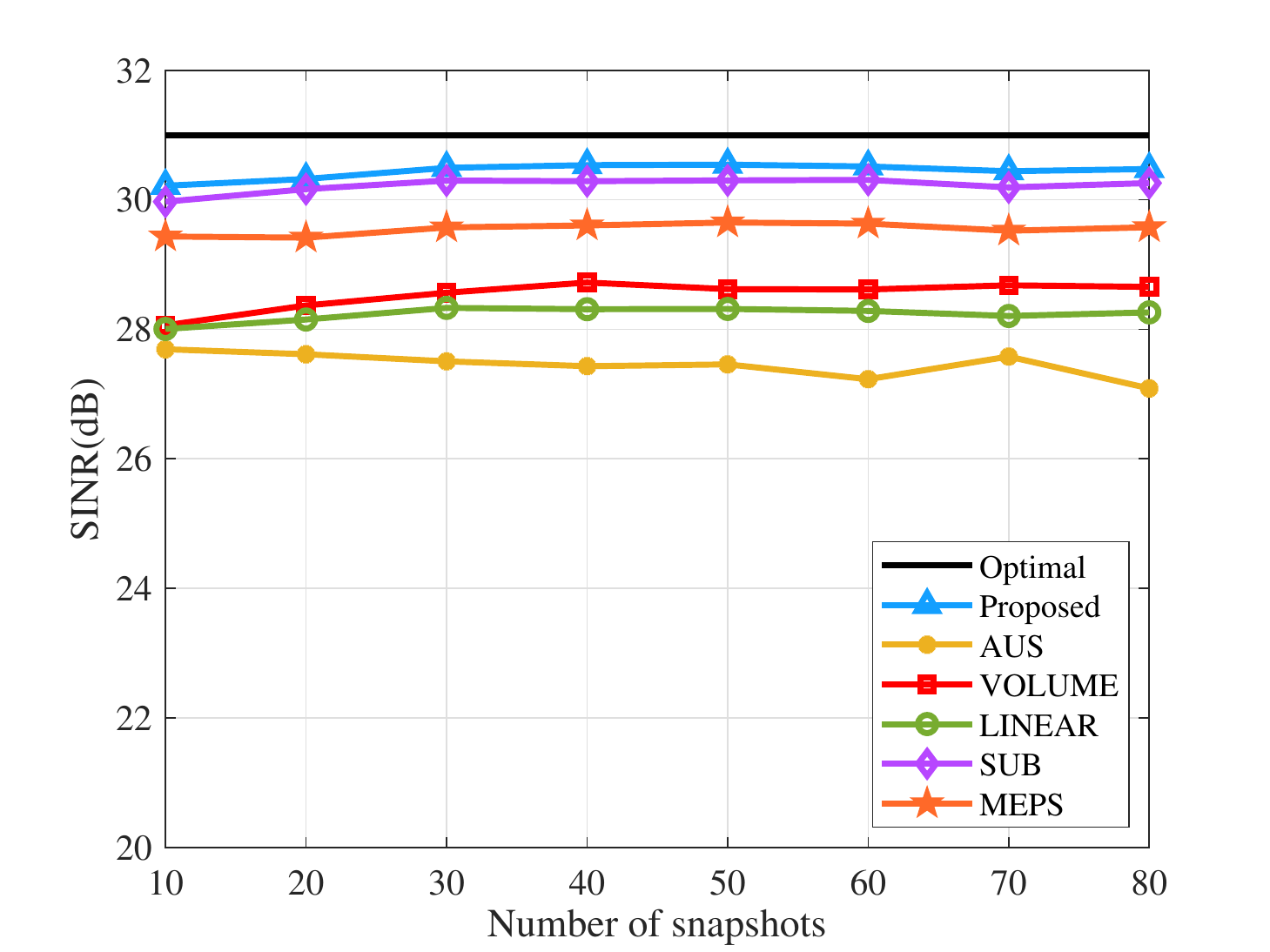}
		\caption{Output SINR versus the number of snapshots in example 1.}
		\label{ep13}
	\end{figure}
	
	The performance of the output SINR versus the input SNR is shown in Fig.~\ref{ep11}. To better distinguish the differences among each algorithm, Fig.~\ref{ep12} demonstrates the deviation between the optimal and other beamformers. The two figures show that the proposed algorithm has a better performance compared to other beamformers in the case of random DOA mismatch. Although at the low SNR, the MEPS performs a little better than the proposed, the proposed algorithm has an excellent performance in most cases. The performance of SUB gets decline when the SNR is lower. It may be caused by the projection operation since the subspace is constructed based on the eigenvalue decomposition, which is inaccurate when the SNR is low. Fig.~\ref{ep13} depicts the output SINR when the input SNR is fixed at $20$ dB and the number of snapshots is varied. It can be found that the performances of all the algorithms fluctuate slightly when the snapshot number changes, while the proposed algorithm always has the best performance and is more robust.
	
	To make it more general, we test the proposed algorithm when the desired signal angle is close to the interference direction. In this simulation, the directions of desired signal and interference are set as $\theta_{0}=5^{\circ}$, $\theta_{1}=-5^{\circ}$, and $\theta_{2}=15^{\circ}$, respectively. Other parameters are the same as previous simulations, and the results are shown in Fig. \ref{xz}. The proposed algorithm has excellent performance, but the performance of the compared algorithms declines obviously, demonstrating that the proposed algorithm has better robustness.  
	\begin{table}
		\begin{figure}[H]
			\includegraphics[width=0.5\textwidth]{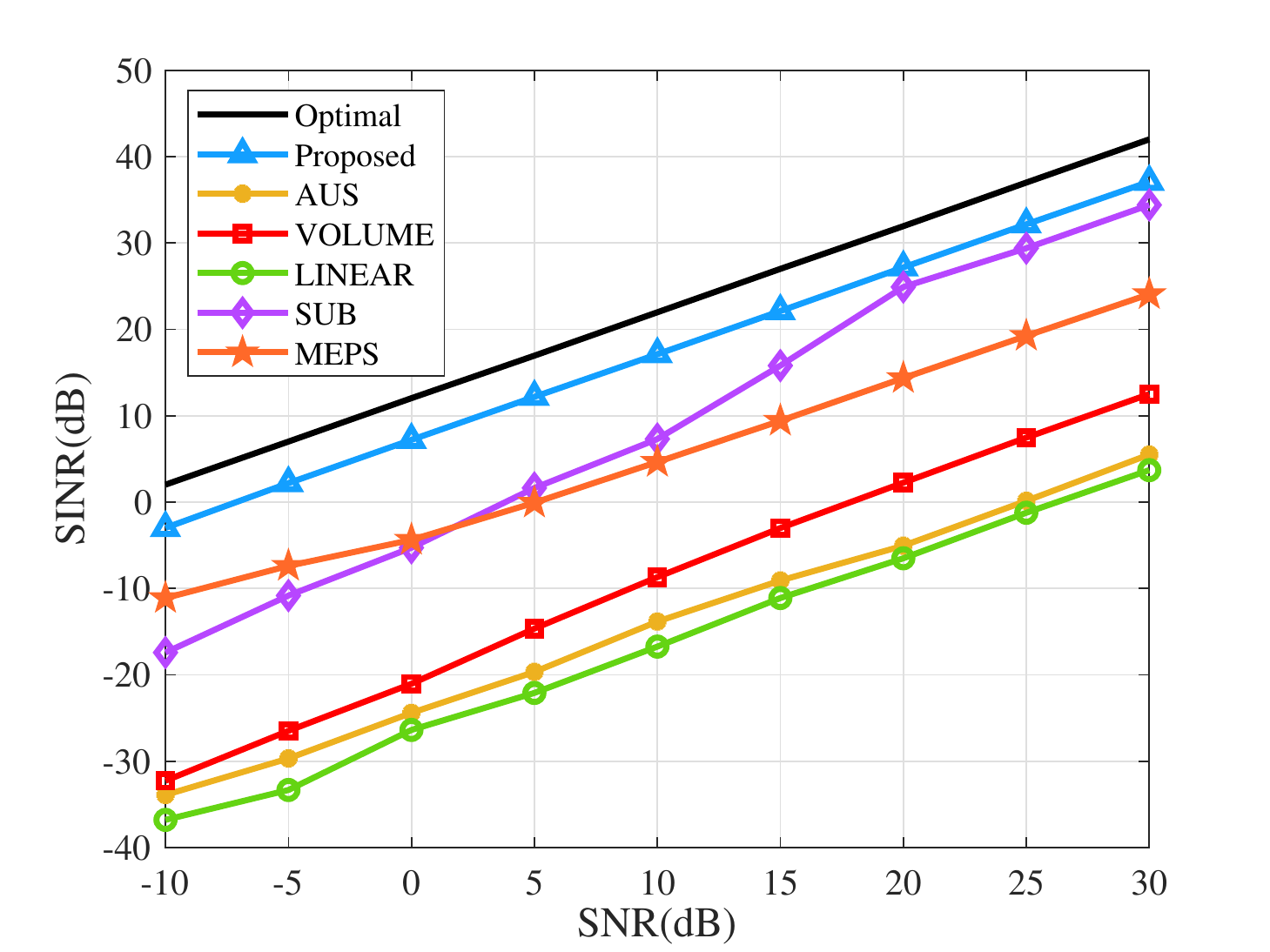}
			\caption{Output SINR versus input SNR for closer angles.}
			\label{xz}
		\end{figure}
	\end{table}

	\subsection{Example 3: Mismatch  due to Gain and Phase Perturbations}
	In this example, the influence of gain and phase perturbations on array output is considered. The gain and phase mismatches on the $m$-th sensor can be described as
	\begin{equation}
		a_{m}(\theta) = (1+\gamma_{m})e^{j(\pi\mathrm{sin}\theta(m-1)+\delta_{m})},
	\end{equation}
	where $\gamma_{m}\in\mathcal{N}(0,0.05^{2})$ denotes the zero-mean random gain perturbation of  $m$-th sensor, and $\delta_{m}\in\mathcal{N}(0,(0.025\pi)^{2})$ denotes the zero-mean random phase perturbation of $m$-th sensor.
	\begin{figure}[ht]
		\centering 
		\includegraphics[width=0.5\textwidth]{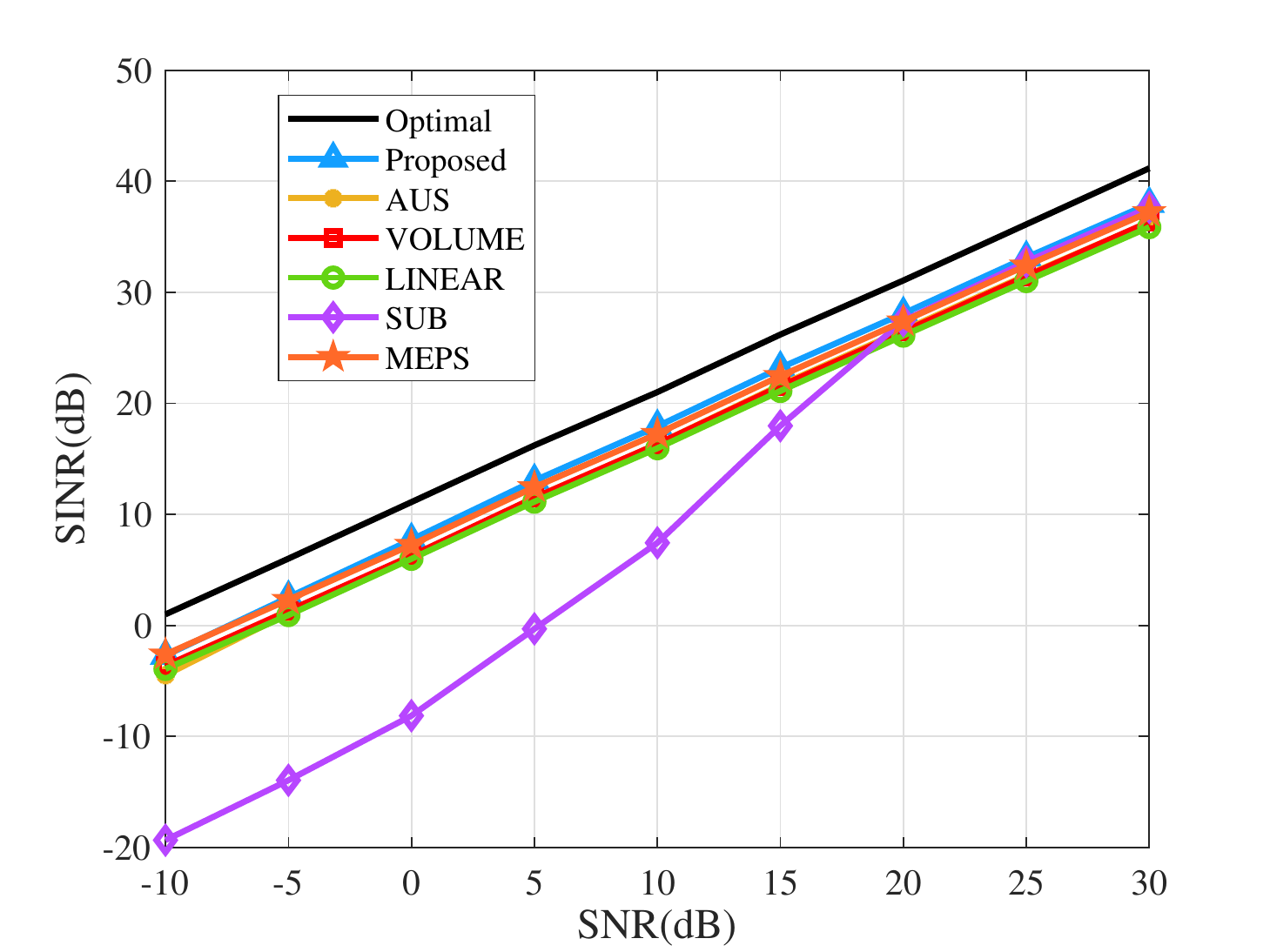}
		\caption{Output SINR versus input SNR in example 2.}
		\label{ep21}
	\end{figure}
	\begin{figure}[ht]
		\centering 
		\includegraphics[width=0.5\textwidth]{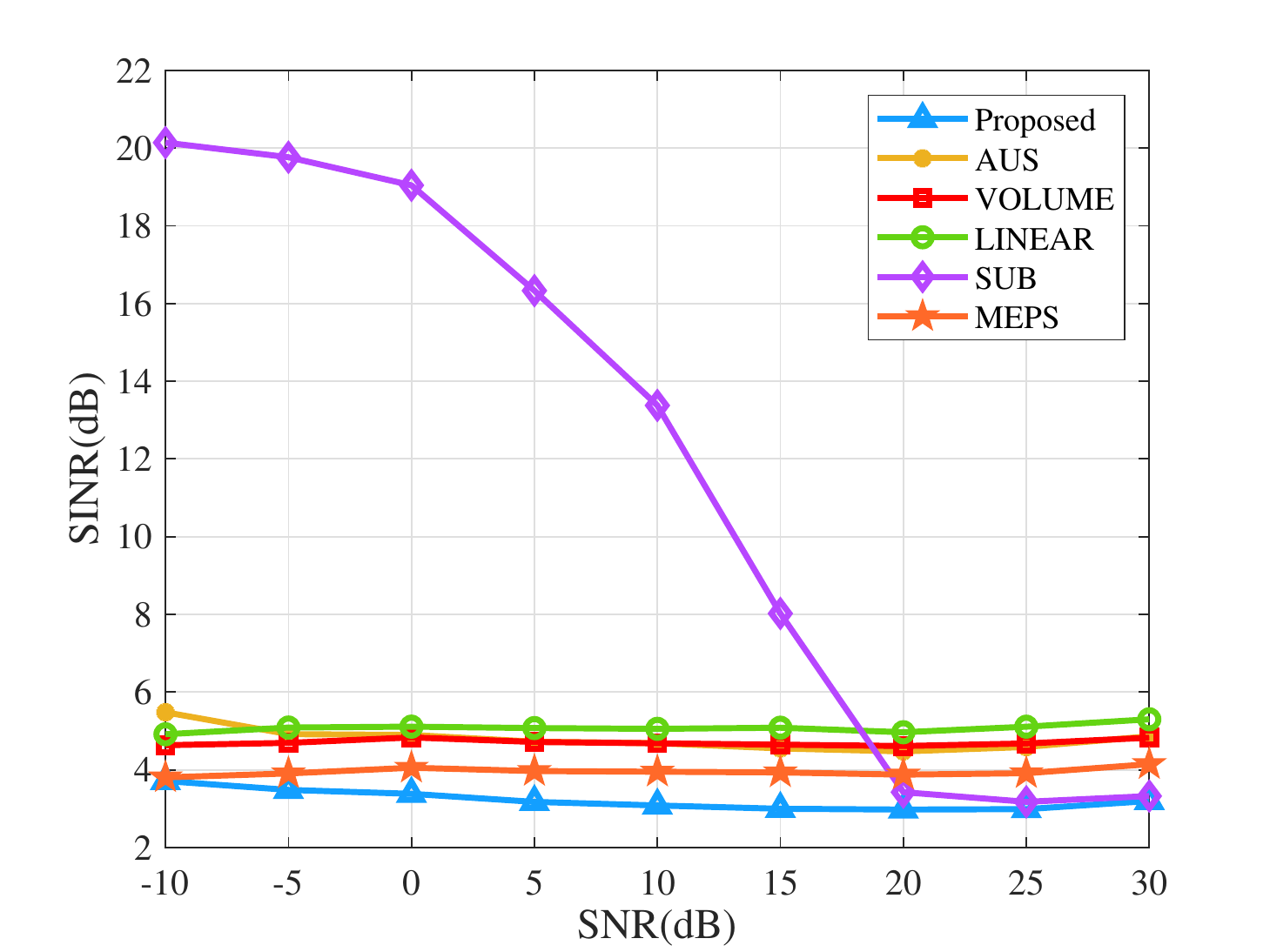}
		\caption{Deviation from optimal SINR versus SNR in example 2.}
		\label{ep22}
	\end{figure}
	\begin{figure}[ht]
		\centering 
		\includegraphics[width=0.5\textwidth]{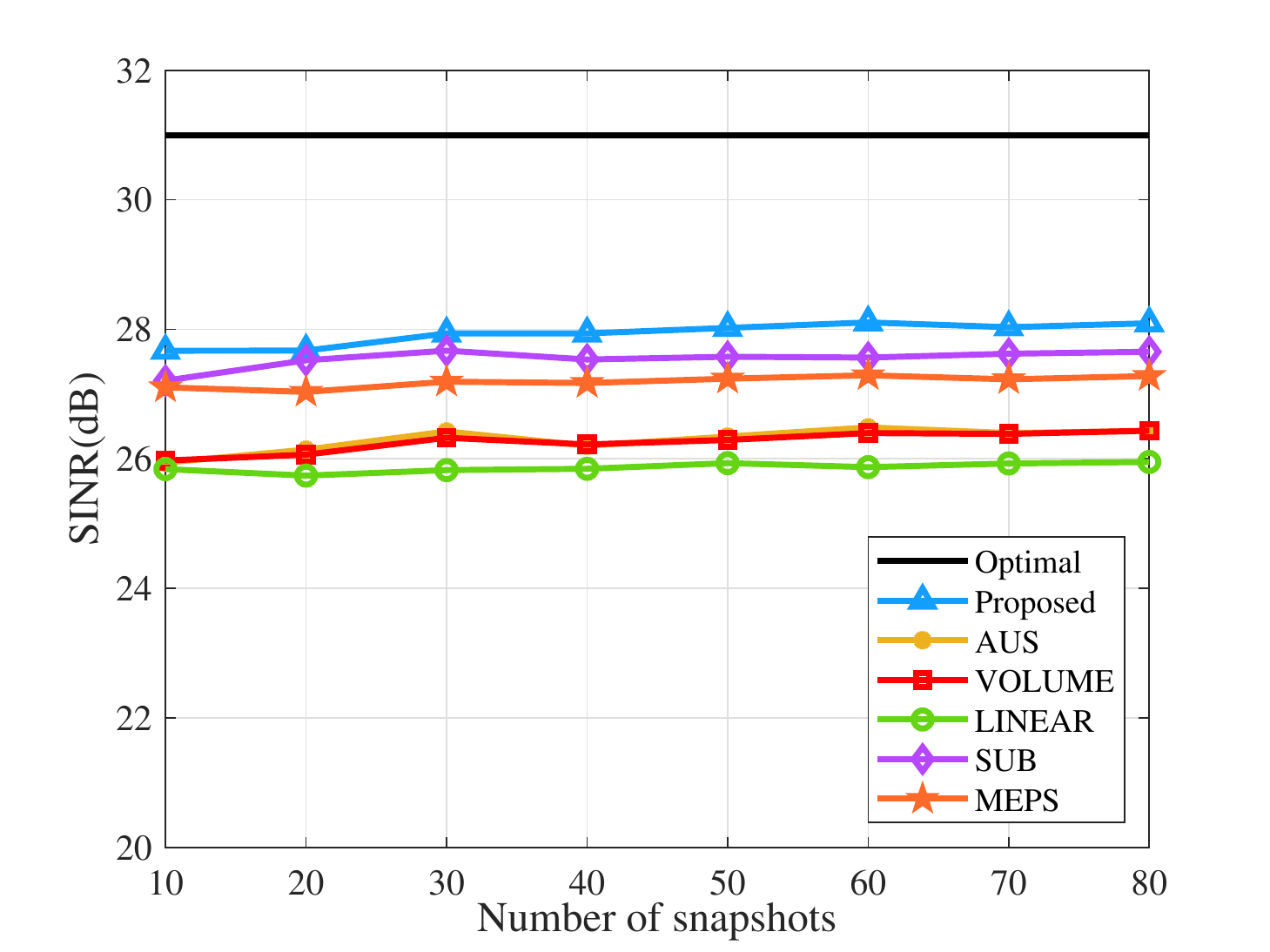}
		\caption{Output SINR versus the number of snapshots in example 2.}
		\label{ep23}
	\end{figure}
	
	Fig.~\ref{ep21} reveals the performance curves versus the input SNR, while Fig.~\ref{ep22} shows the deviation between the optimal method and the tested algorithm, respectively. As they show, even if all the beamformers have great performances, the proposed algorithm gets the best in the case of gain and phase perturbations. The SUB algorithm has excellent performance only when the SNR is higher than $15$ dB. Fig.~\ref{ep22} shows that the proposed algorithm has the performance which is closest to the optimal. The performance versus the number of snapshots is demonstrated in Fig.~\ref{ep23}. The AUS and VOLUME have almost the same performance, while the proposed performs the best regardless of the number of snapshots. Additionally, as the simulation results show, all the algorithms can not achieve the optimal output SINR. It is mainly determined that the optimal SINR is calculated by error-free parameters, while the beamforming algorithms work in a complex environment with errors.
	
	\subsection{Example 4: Mismatch due to Steering Vector Random Error}
	The effect of mismatch due to steering vector random error is investigated in this example. Considering that each actual steering vector is generated by adding a random error vector to the nominal steering vector as
	\begin{equation}
		\overline{\boldsymbol{a}}_{i} = \boldsymbol{a}_{i} + \boldsymbol{\xi}_{i},
	\end{equation}
	where $\boldsymbol{\xi}_{i}$ is the random error corresponding to $\boldsymbol{a}_{i}$, and can be expressed as
	\begin{equation}
		\boldsymbol{\xi}_{i} = \frac{\rho_{i}}{\sqrt{M}}[e^{j\phi_{0}^{i}}, e^{j\phi_{2}^{i}}, \dots, e^{j\phi_{M-1}^{i}}]^{\mathrm{T}},
	\end{equation} 
	where the Euclidean norm $\rho_{i}$ follows a uniform distribution in the interval [0,$\sqrt{0.3}$], and the phases $\phi_{m}^{i}, m=0, 1, \dots, M-1$ are uniformly distributed in [$0$, $2\pi$) and are independent with each other.
	\begin{figure}[t]
		\centering 
		\includegraphics[width=0.5\textwidth]{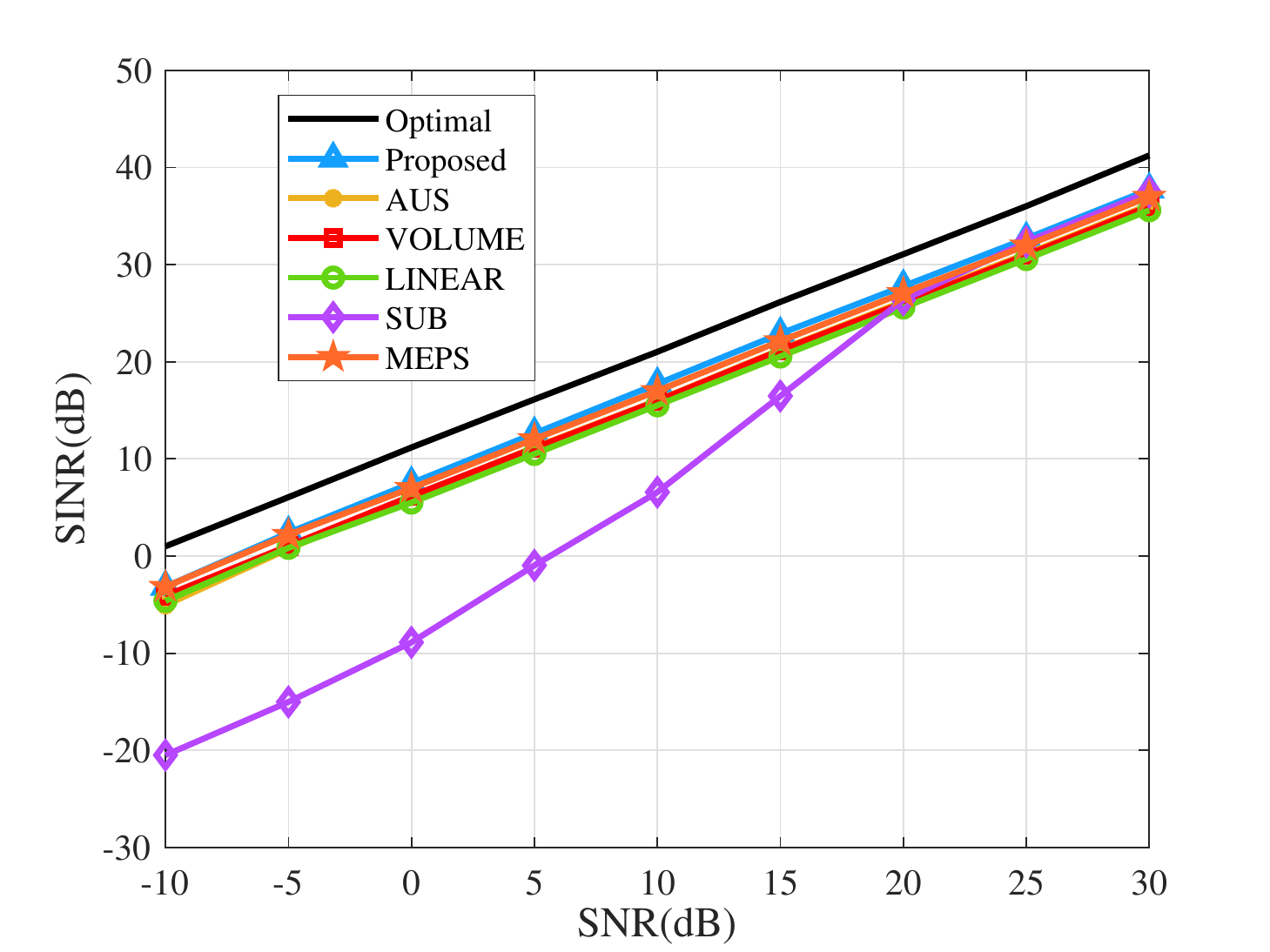}
		\caption{Output SINR versus input SNR in example 3.}
		\label{ep31}
	\end{figure}
	\begin{figure}[t]
		\centering 
		\includegraphics[width=0.5\textwidth]{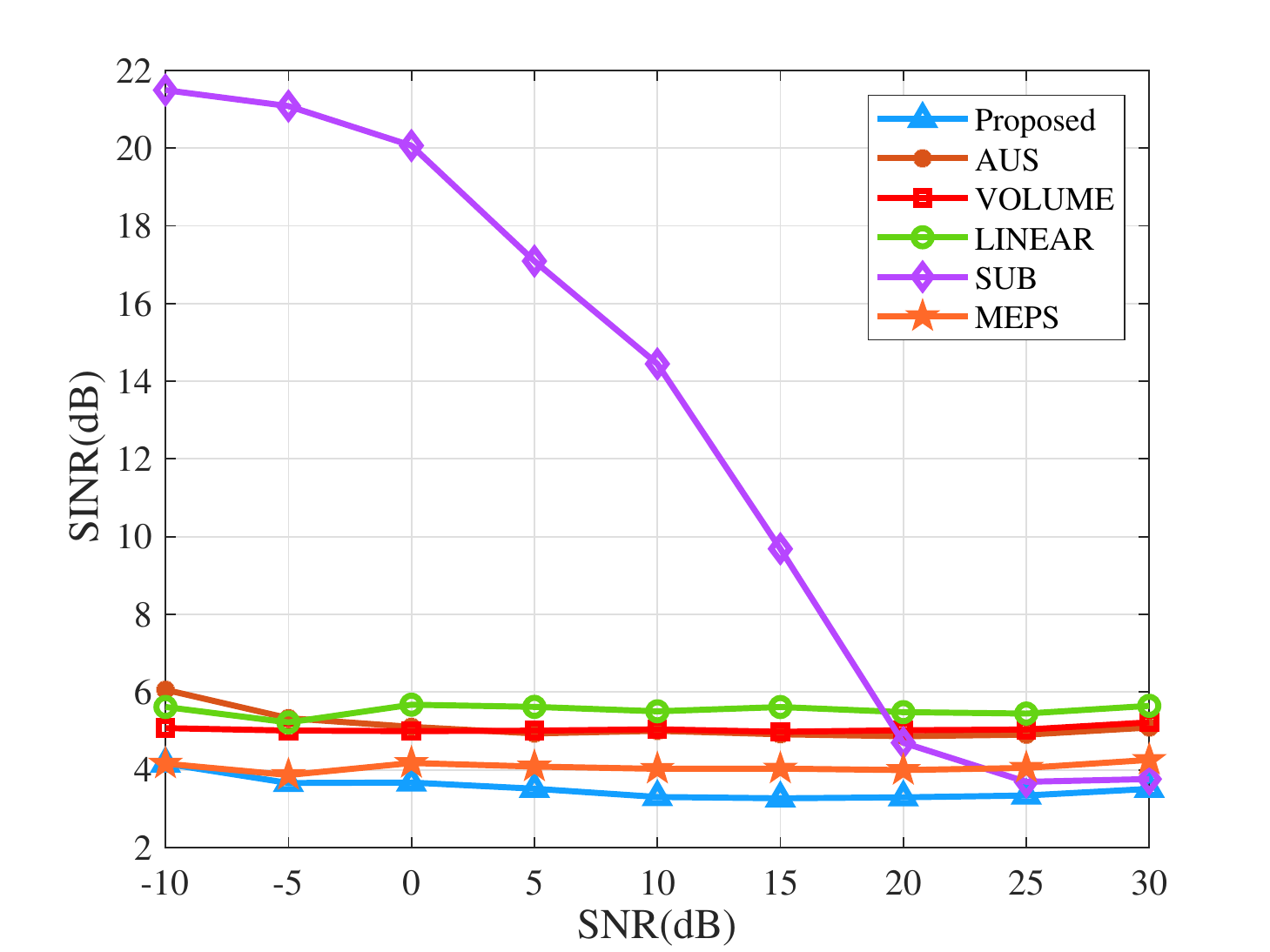}
		\caption{Deviation from optimal SINR versus SNR in example 3.}
		\label{ep32}
	\end{figure}
	\begin{figure}[t]
		\centering 
		\includegraphics[width=0.45\textwidth]{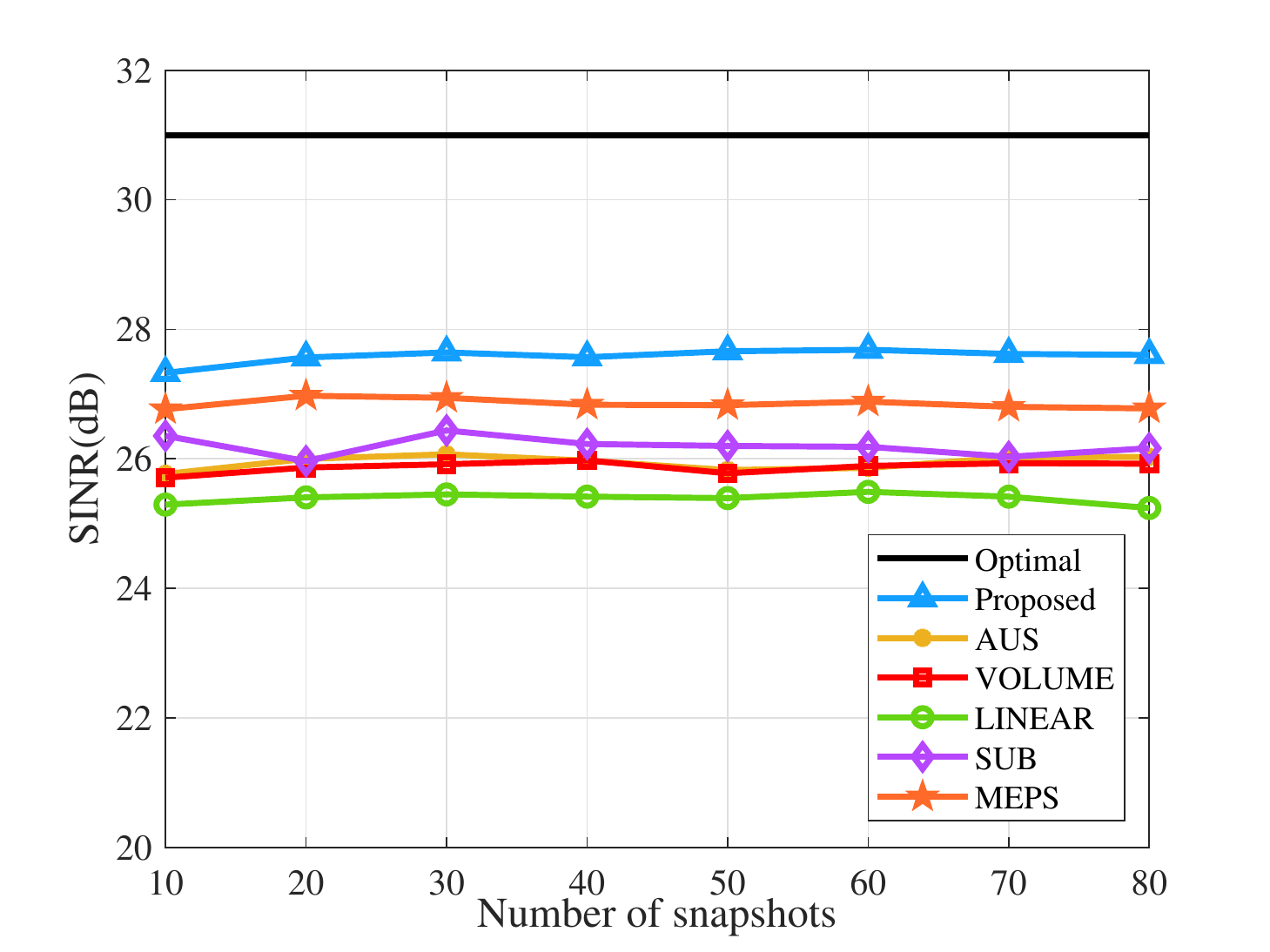}
		\caption{Output SINR versus the number of snapshots in example 3.}
		\label{ep33}
	\end{figure}
	
	Fig.~\ref{ep31} and Fig.~\ref{ep32} display the performance of the tested algorithms versus the input SNR. It can be found that in the case of steering vector random error, even if the MEPS and other algorithms can effectively reduce the error, the proposed algorithm demonstrates an obvious improvement compared to other beamformers. The AUS performs with a good effect which is the same as VOLUME. Fig.~\ref{ep33} displays the output SINR versus the number of snapshots, which shows the proposed algorithm has excellent robustness.
	\subsection{Experimental Data}
\begin{figure}[t]
		\centering 
		\includegraphics[width=0.5\textwidth]{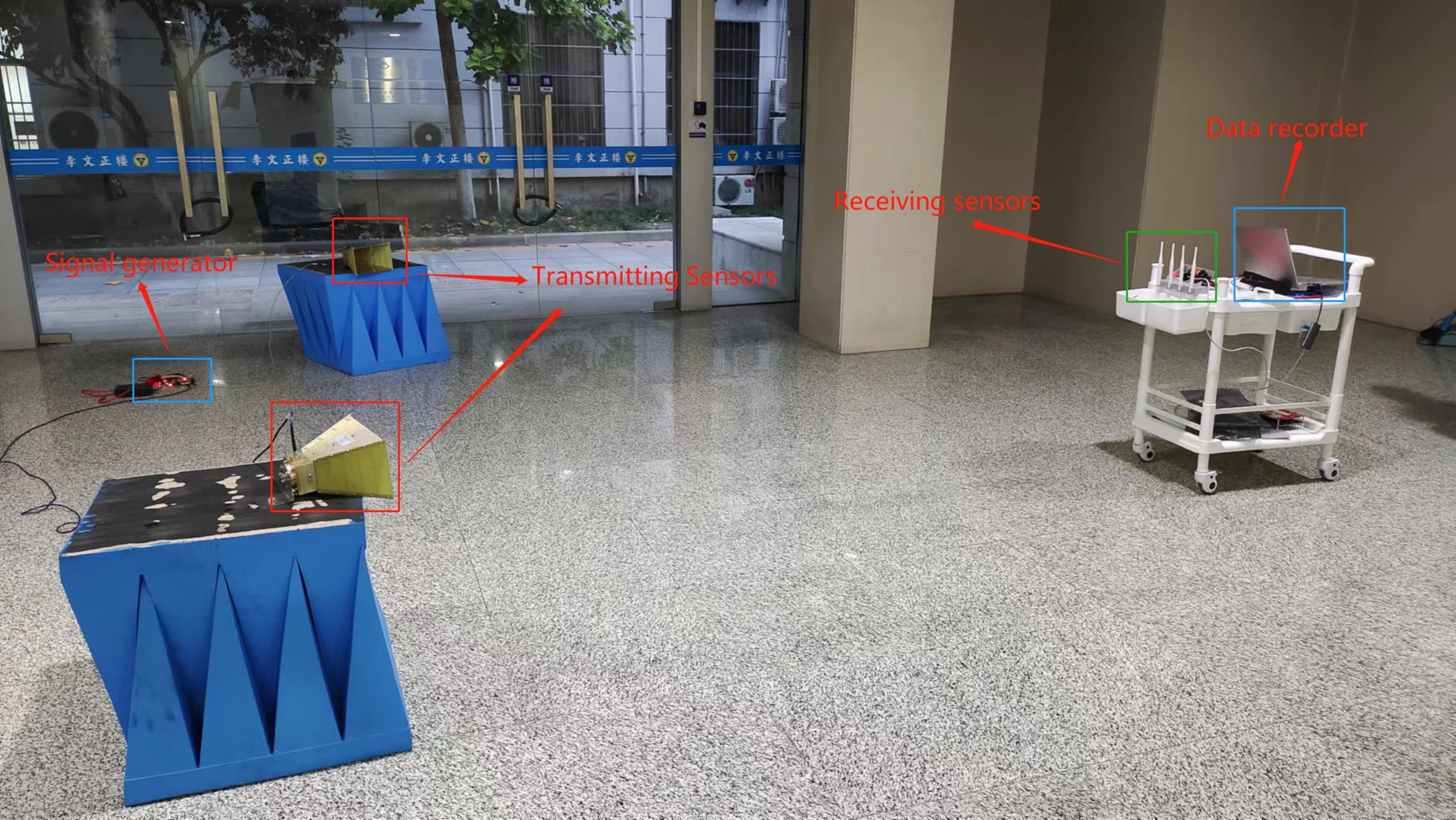}
		\caption{The experimental scene.}
		\label{exper}
	\end{figure}

	In this section, an experiment is carried out by an S-band ULA system. The system consists of $4$ received sensors and $2$ transmitting sensors, where the transmitting sensors are both about $4$ meters away from the receiving sensors, and the spacing between the two transmitting sensors is about $4.2$ meters. The spacing of adjacent receiving sensors is half-wavelength. Signals are generated by a field-programmable gate array (FPGA) signal generator and consist of Gaussian white noise with $10$ MHz bandwidth. The received signals are transmitted to the computer through the network cable for subsequent processing. The experimental scene is presented in Fig.~\ref{exper}. 
	
	According to the experimental data, we get the two transmitting sensors' locations at $-54.8^{\circ}$ and $8.1^{\circ}$ of the receiving sensors by using the Capon spectrum, respectively. The transmitting sensor at $-54.8^{\circ}$ is chosen as the interference, and the other is chosen as the target. The SNR and INR of the received signal are both $5$ dB. 
	\begin{figure}[t]
		\centering 
		\includegraphics[width=0.5\textwidth]{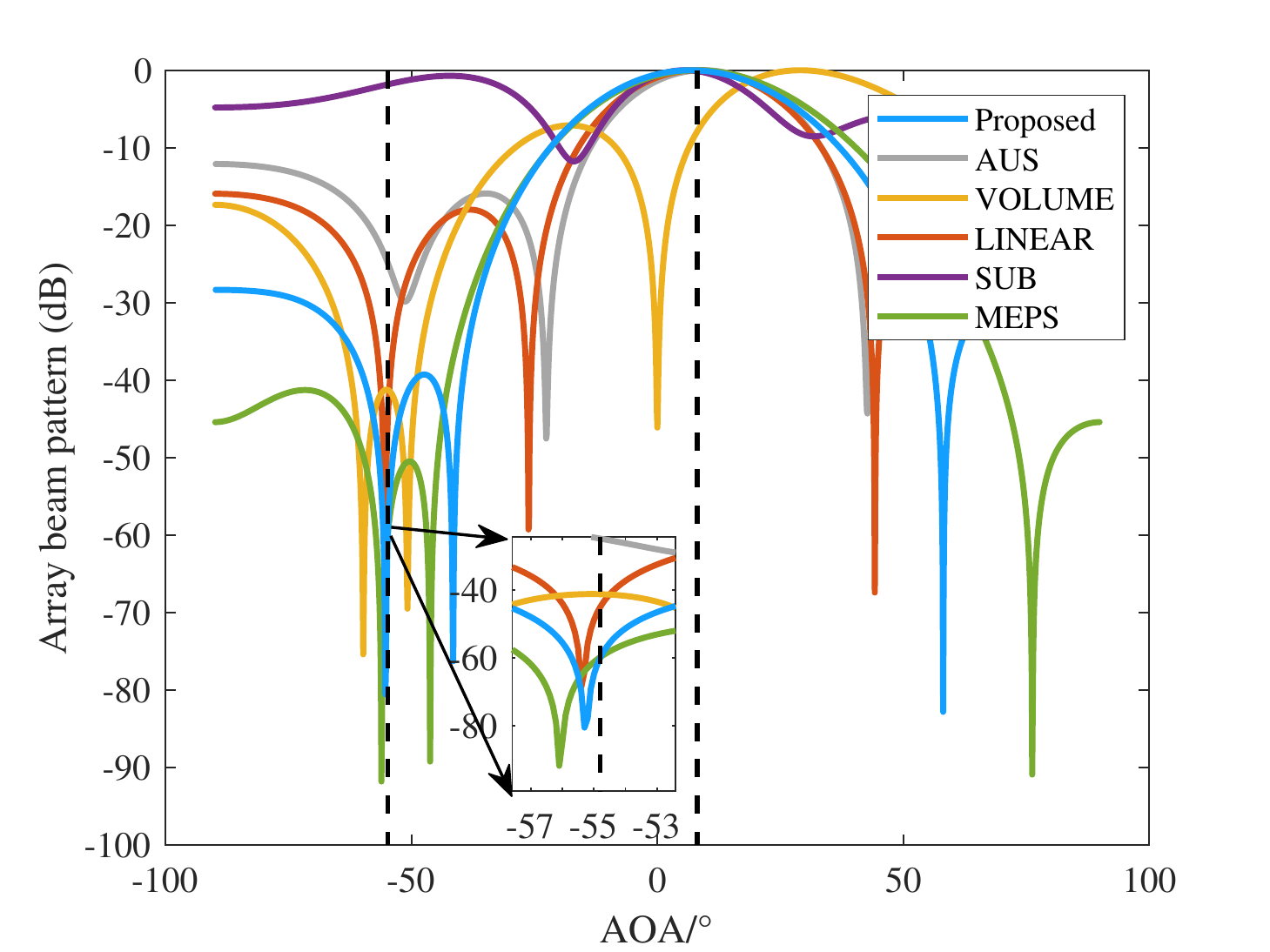}
		\caption{The beampattern for the different algorithms with experimental data.}
		\label{shice}
	\end{figure}
	
	The performance of the proposed algorithm is evaluated and compared with AUS, VOLUME, LINEAR, SUB, and MEPS methods. The beampatterns of all the tested algorithms with the experimental data are shown in Fig.~\ref{shice}. We can find that almost all the algorithms can suppress the interference efficiently except the SUB method, which is mainly caused by the low SNR, and is consistent with the previous simulation results. The proposed algorithm forms a deeper null around the location of the interference, which is the same as MEPS algorithm but more precisely than MEPS, since the null is closer to the direction of the interference. Furthermore, the proposed algorithm maintains the desired signal well, it only forms little attenuation in the direction of the desired signal, while the VOLUME method forms a nearly $10$ dB attenuation of the desired signal. 
 
	\section{Conclusion}\label{secconc}
	In this work, a novel adaptive beamforming algorithm based on the IPNCM reconstruction is proposed. A projection matrix is constructed to remove the desired signal from the received data to reconstruct the IPNCM accurately. To reduce the algorithm complexity, the Gauss-Legendre quadrature is introduced. Based on the reconstructed covariance matrix, the presumed steering vector of the signal is corrected by maximizing the array output power. The computational complexity of the proposed algorithm is $\mathcal{O}(M^{3.5})$, which is less than most robust adaptive beamformers. The simulation results show that compared to other beamformers, the proposed beamformer can achieve excellent performance with less computation and is always close to the optimal. Meanwhile, the experimental data shows the proposed algorithm performs better in a practical environment. Future work will explore novel reconstruction methods of IPNCM to improve the robustness of the beamformer further and reduce the computational complexity. 
	
\bibliographystyle{IEEEtran}
\bibliography{IEEEabrv,References}

\vspace{-1cm} 
\begin{IEEEbiography}[{\includegraphics[width=1in,height=1.25in,clip,keepaspectratio]{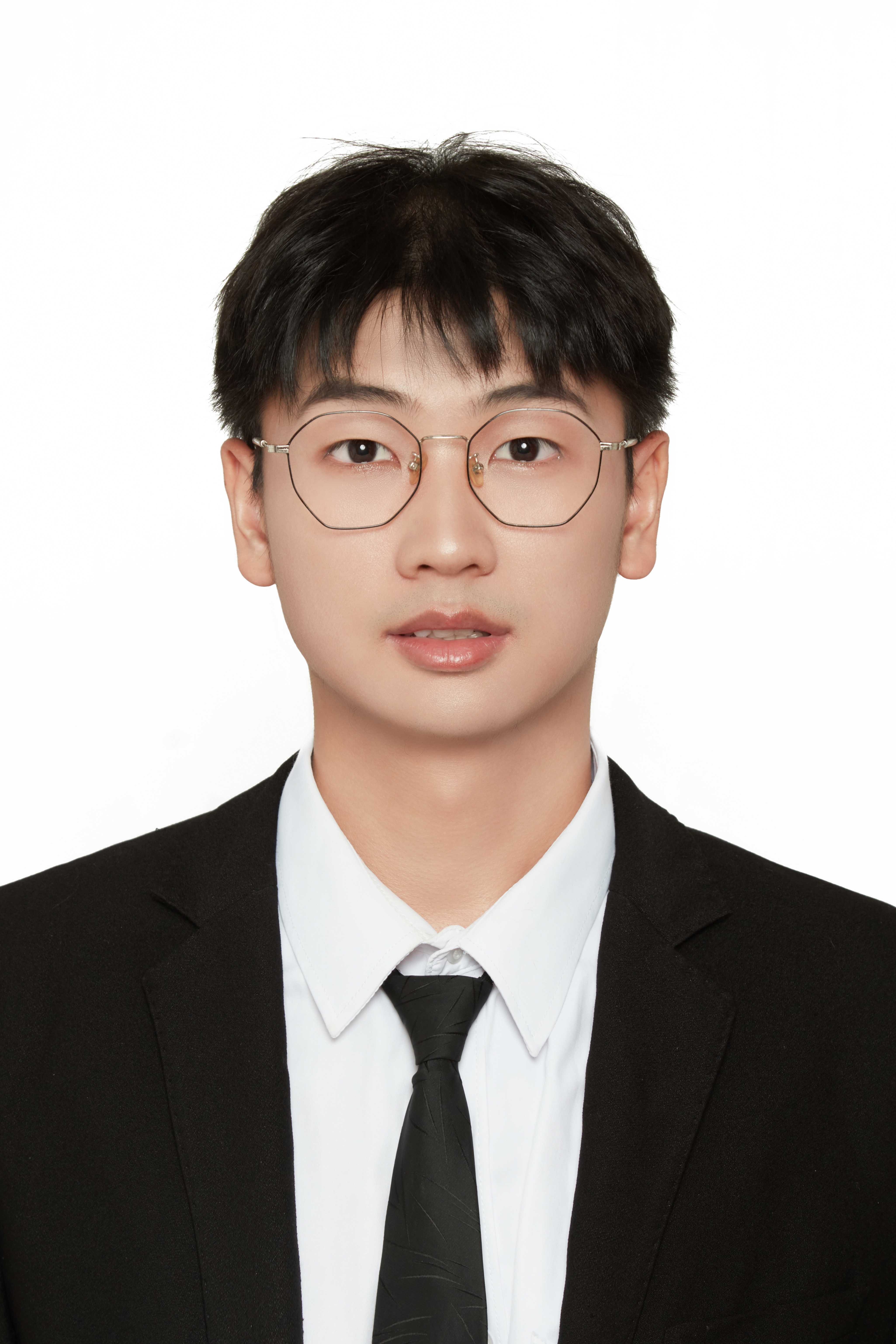}}]{Tao Luo} was born in Hunan, China in 1999. He received the B.E. degree in 2021 from the School of Electronic and Optical Engineering, Nanjing University of Science and Technology, China. He is currently pursuing the Ph.D. degree in the School of Information Science and Engineering, Southeast University, China. His research interests include radar signal processing and millimeter wave communication.
\end{IEEEbiography}
\vspace{-1cm} 
\begin{IEEEbiography}[{\includegraphics[width=1in,height=1.25in,clip,keepaspectratio]{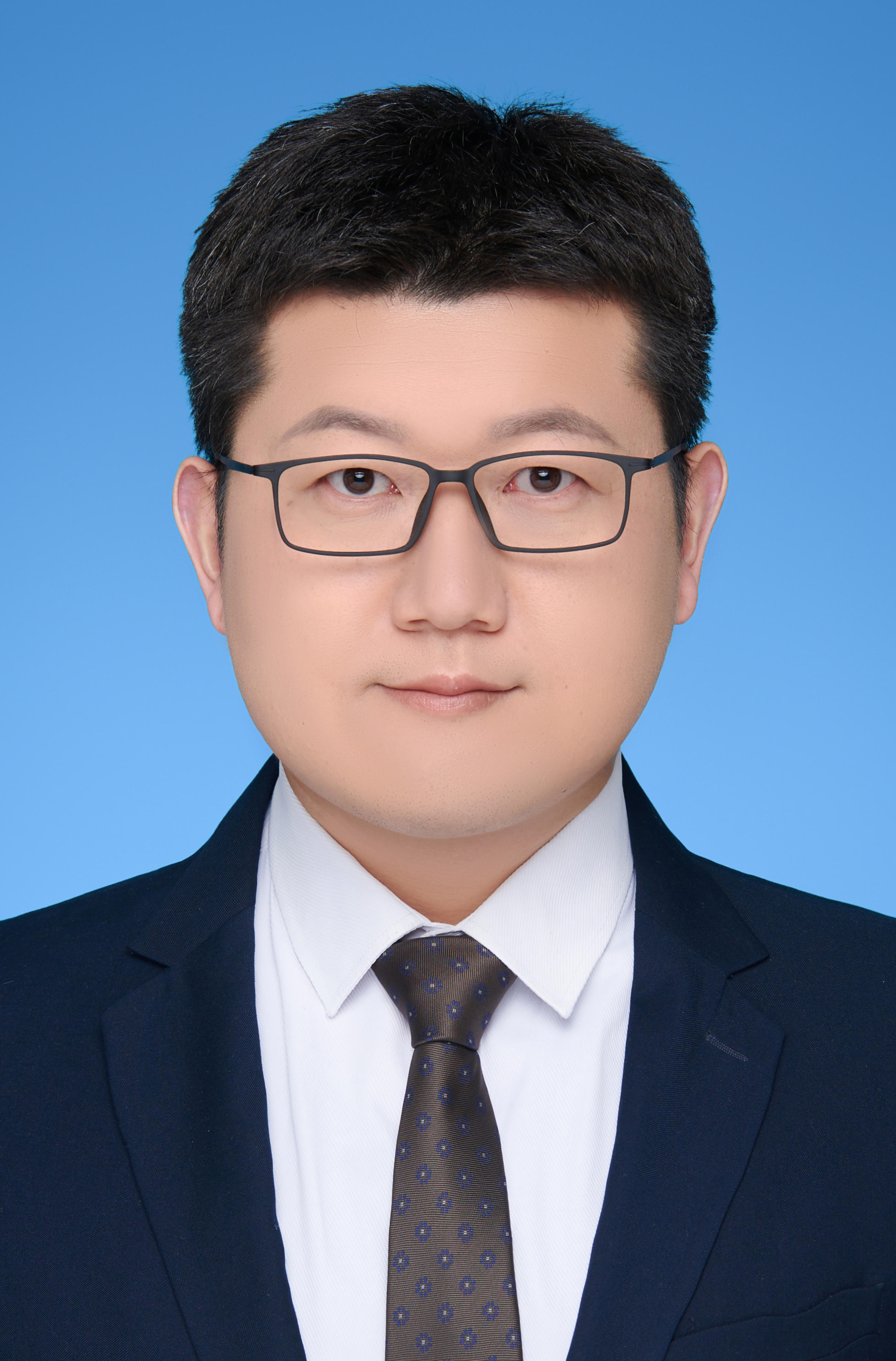}}]{Peng Chen (Senior Member, IEEE)} was born in Jiangsu, China, in 1989. He received the B.E. and Ph.D. degrees from the School of Information Science and Engineering, Southeast University, Nanjing, China, in 2011 and 2017 respectively. From March 2015 to April 2016, he was a Visiting Scholar with the Department of Electrical Engineering, Columbia University, New York, NY, USA. He is currently an Associate Professor with the State Key Laboratory of Millimeter Waves, Southeast University. His research interests include target localization, super-resolution reconstruction, and array signal processing.

He is a Jiangsu Province Outstanding Young Scientist. He has served as an IEEE ICCC Session Chair, and won the Best Presentation Award in 2022 (IEEE ICCC). He was invited as a keynote speaker at the IEEE ICET in 2022. He was recognized as an exemplary reviewer for IEEE WCL in 2021, and won the Best Paper Award at IEEE ICCCCEE in 2017.
\end{IEEEbiography}
\vspace{-1cm} 
\begin{IEEEbiography}[{\includegraphics[width=1in,height=1.25in,clip,keepaspectratio]{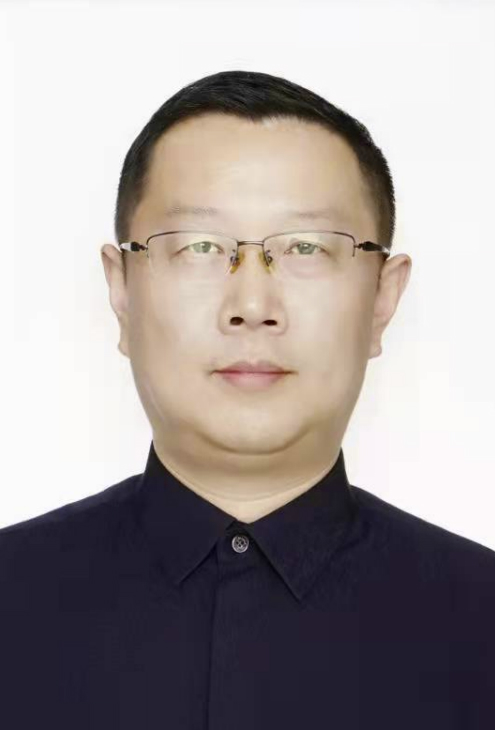}}]{Zhenxin Cao (Member, IEEE)} was born in May 1976. He received the M.S. degree from Nanjing University of Aeronautics and Astronautics, Nanjing, China, in 2002 and the Ph.D. degree from the School of Information Science and Engineering, Southeast University, Nanjing, China, in 2005. From 2012 to 2013, he was a Visiting Scholar with North Carolina State University. Since 2005, he has been with the State Key Laboratory of Millimeter Waves, Southeast University, where he is currently a Professor. His research interests include antenna theory and application.
\end{IEEEbiography}
\vspace{-1cm} 
\begin{IEEEbiography}[{\includegraphics[width=1in,height=1.25in,clip,keepaspectratio]{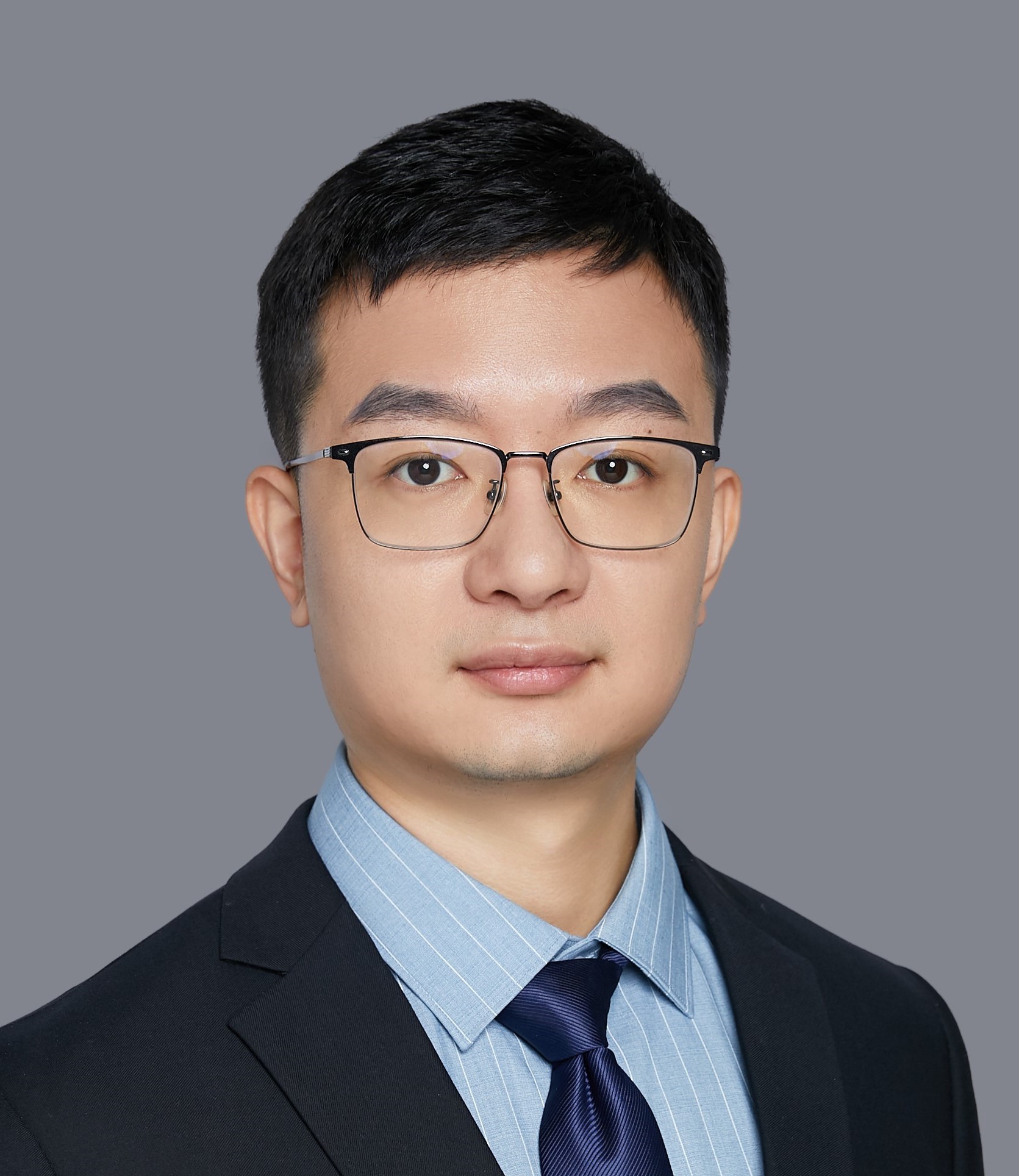}}]{Le Zheng (Senior Member, IEEE)} received the B.Eng. degree from Northwestern Polytechnical University (NWPU), Xi’an, China, in 2009 and Ph.D degree from Beijing Institute of Technology (BIT), Beijing, China in 2015, respectively. He has previously held academic positions in the Electrical Engineering Department of Columbia University, New York, U.S., first as a Visiting Researcher from 2013 to 2014 and then as a Postdoc Research Fellow from 2015 to 2017. From 2018 to 2022, he worked at Aptiv (formerly Delphi), Los Angeles, as a Principal Radar Systems Engineer, leading projects on the next-generation automotive radar products. Since 2022, he has been a Full Professor with the School of Information and Electronics, BIT. His research interests lie in the general areas of radar, statistical signal processing, wireless communication, and high-performance hardware, and in particular in the area of automotive radar and integrated sensing and communications (ISAC).
\end{IEEEbiography}
\vspace{-1cm} 
\begin{IEEEbiography}[{\includegraphics[width=1in,height=1.25in,clip,keepaspectratio]{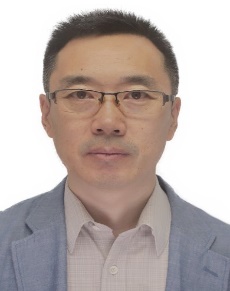}}]{Zongxin Wang (Member, IEEE)} was born in Yantai, Shandong province, China in 1970. He received the B.S. degree in microelectronics engineering from the University of Hunan, China, in1993, the M.S. degree from the institute of mechanical science, China, in 2001 and the Ph.D. degree from the School of Information Science and Engineering, Southeast University, China, in 2006.

He is currently an Associate Professor with the State Key Laboratory of Millimeter Waves, Southeast University. His research interests include PCB antenna array, slotted waveguide array and antenna stealth technique.
\end{IEEEbiography}

\end{document}